%% file: erasures_extended.tex
\documentclass[11pt, letterpaper, onecolumn]{IEEEtran}

\input{preamble}

\newcommand{\eu}{\epsilon_{\rmu}} 
\newcommand{\et}{\epsilon_{\rmt}} 
\newcommand{\ee}{\epsilon_{\rme}} 
\newcommand{\eun}{\epsilon_{\rmu,n}} 
\newcommand{\etn}{\epsilon_{\rmt,n}} 
 
\newcommand{\eeu}{\lambda_{\rmu}} 
\newcommand{\eet}{\lambda_{\rmt}} 
\newcommand{\eee}{\lambda_{\rme}} 
\usepackage[ colorlinks = true,
             linkcolor = blue,
             urlcolor  = blue,
             citecolor = red,
             anchorcolor = green,
]{hyperref}

\begin{document}
\flushbottom

\title{Fixed Error Probability   Asymptotics For    Erasure and List Decoding}

\author{Vincent Y.~F.\ Tan$\,\,\, $  and $\,\,$ Pierre Moulin \thanks{V.~.Y.~F. Tan is with the Department of Electrical and Computer Engineering (ECE) and the Department of Mathematics, National University of Singapore (Email: \url{vtan@nus.edu.sg}).    P.~Moulin is with the Department of Electrical and Computer Engineering (ECE), University of Illinois at Urbana-Champaign (Email: \url{moulin@ifp.uiuc.edu}).   } }


\maketitle

\begin{abstract}
We  derive the optimum second-order coding rates, known as second-order capacities, for erasure and list decoding. For erasure decoding for discrete memoryless channels, we show that second-order capacity is  $\sqrt{V}\Phi^{-1}(\et)$ where $V$ is the channel dispersion and $\et$ is the total error probability, i.e., the sum of the erasure and undetected errors.  This total error probability, as are other error probabilities in this paper, is fixed at a non-zero constant. We show numerically that the expected rate at finite blocklength for erasures decoding can exceed the finite blocklength channel coding rate. We also show that the analogous result also holds for lossless source coding with decoder side information, i.e., Slepian-Wolf coding. For list decoding, we consider list codes of deterministic sizes  that scale as $\exp(\sqrt{n} \, l)$ and show that the corresponding second-order capacity is $l+\sqrt{V}\Phi^{-1}(\epsilon)$, where $\epsilon$ is the permissible error probability, i.e., the probability that the true message is not in the list. We also consider lists of polynomial size $n^\alpha$ and derive bounds on the third-order coding rate in terms of the order of the polynomial $\alpha$.  These bounds are tight for symmetric and singular channels. The direct parts of the coding theorems leverage on    the simple threshold decoder and converses are proved  using  variants  of the hypothesis testing converse. 
\end{abstract}

\section{Introduction}
In many communication scenarios, it is advantageous to allow the decoder to have the option of either not deciding at all or putting out more than one estimate of the   message. These are respectively known as  {\em erasure} and {\em list} decoding respectively and have been studied extensively in the information theory literature~\cite{Forney68,sgb, tel94, Blinovsky, Moulin09, merhav08,merhav13}. The erasure and list options generally allow  for   smaller undetected error probabilities so these options are useful in practice. 

In this paper, we revisit the problem or erasure and list decoding  from the viewpoint of second- and third-order asymptotics. The study of second-order   asymptotics for fixed (non-vanishing) error probability  was first done by Strassen~\cite[Thm.~1.2]{Strassen}  who showed for a well-behaved discrete memoryless channel  (DMCs) $W$ that the maximum number of codewords at (average or maximum) error probability $\epsilon$, namely $M^*(W^n,\epsilon)$, satisfies 
\begin{equation}
\log M^*(W^n,\epsilon)=nC+ \sqrt{nV}\Phi^{-1}(\epsilon)+ O(\log n), \label{eqn:strassen}
\end{equation}
where $C$ and $V$ are respectively the capacity and the dispersion of $W$ and $\Phi^{-1}(\cdot)$ is the inverse of the standard Gaussian cumulative distribution function (cdf). Also see Kemperman~\cite[Thm.~11.3]{Kemperman} for the corresponding result for constant composition codes.  This line of work has been revisited by numerous authors recently  and they considered various other channel models such as the additive white Gaussian (AWGN) channel~\cite{PPV10,Hayashi09}, bounds on the third-order (logarithmic) term  \cite{mou13b, TomTan12, altug13,TanTom13} and extensions to multi-terminal problems such as the two-encoder Slepian-Wolf problem and the multiple-access channel~\cite{TK14}.

\subsection{Main Contributions}
In this paper, for erasure decoding, we consider constant undetected and total (sum of undetected and erasure)  error probabilities (numbers between $0$ and $1$) and we obtain the analogue of the second-order $\sqrt{n}$ term in \eqref{eqn:strassen}.  We show that  the coefficient of the second-order term, termed the {\em second-order capacity}, is $\sqrt{V}\Phi^{-1}(\et)$ where $\et$ is the total error probability. The second-order capacity is thus completely independent  of the undetected error probability. We then compute the expected rate at finite blocklength allowing erasures and show that it can exceed the finite blocklength rate without the erasure option, i.e., usual channel coding. We show that these results carry over in a straightforward manner to the problem of lossless source coding with side information, i.e., the Slepian-Wolf problem~\cite{sw73} which was previously studied in the for the case without the erasure option~\cite{TK14}. 

For list decoding, we consider lists of deterministic size of order $\exp(\sqrt{n}\, l)$ and show that the second-order capacity is  $l+\sqrt{V}\Phi^{-1}(\epsilon)$, where $\epsilon$ is the permissible error probability.  We also consider lists of polynomial size $n^\alpha$ and demonstrate bounds on the third-order term in terms of the degree of the polynomial $\alpha$. These bounds turn out to tight for channels that are symmetric and singular--a canonical example being the binary erasure channel~\cite{altug13}.  To the best of the authors' knowledge, this is the first time that lists of size other than constant or exponential have been considered in the literature. Practically, the advantage of smaller lists is that they result in lower search complexity for the true message within the decoded list. 

\subsection{Related Work}
Previously, the study of erasure and list decoding has been primarily from the error exponents perspective. We summarize some existing works here.  Forney~\cite{Forney68} derived   optimal decision rules   by generalizing the Neyman-Pearson lemma and also proved exponential upper bounds for the error probabilities using Gallager's Chernoff bounding techniques~\cite{gallagerIT}. Shannon-Gallager-Berlekamp~\cite{sgb} proved exponential lower bounds for the error probabilities and also considered lists of exponential size $\exp(nl)$. They showed that sphere packing error exponent  (evaluated at the code rate minus $l$) is an upper bound on the reliability function~\cite[Ex.~10.28]{Csi97}. Bounds for the error probabilities were derived by Telatar~\cite{tel94} using a general decoder parametrized by an asymmetric relation $\prec$ which is a function of the channel law.   Blinovsky~\cite{Blinovsky} studied the exponents of the  list decoding problem at low (and even $0$) rate.   Csisz\'ar-K\"orner~\cite[Thm.~10.11]{Csi97} present  exponential upper bounds for universally attainable erasure decoding using the method of types. Moulin~\cite{Moulin09} generalized the treatment there and presented improved error exponents under some conditions.  Recently, Merhav also considered alternative methods of analysis~\cite{merhav08} and   expurgated exponents for these problems~\cite{merhav13}. The same author also derived erasure and list exponents for the Slepian-Wolf problem~\cite{merhav13a}.

Another related line of work concerns  constant error probability non-asymptotic fundamental limits of   channel coding with various forms of feedback. Polyanskiy-Poor-Verd\'u~\cite{PPV11b} studied various {\em incremental redundancy} schemes and derived performance  bounds under   receiver- and transmitter-confirmation. In  incremental redundancy systems/schemes, one is allowed to transmit a sequence of coded symbols with numerous opportunities for confirmation before the codeword is resent if necessary.  In contrast   in our study of  channel coding with the erasure option in Sec.~\ref{sec:exp_rate}, we analyze  the expected performance of the  easily-implementable  Forney-style~\cite{Forney68}  {\em single-codeword repetition} scheme  that allows  for confirmation (or erasure) at the end of a complete  codeword block and repeats the same message  until the erasure event no longer occurs.   We compare a quantity termed the {\em expected rate} and to the ordinary channel coding rate. Inspired by  \cite{PPV11b}, Chen {\em et al.}~\cite{chen13} derived non-asymptotic bounds for feedback systems using incremental redundancy  with noiseless transmitter confirmation.  Williamson-Chen-Wesel~\cite{Will13} also improved on the bounds in~\cite{PPV11b} for variable-length feedback coding.
 
\subsection{Structure of Paper} \label{sec:structure}
This paper is structured as follows: In Sec.~\ref{sec:prob_set}, we set the stage by introducing our notation and defining relevant quantities such as the second-order capacity. In Sec.~\ref{sec:main_res}, we state our main results for channel coding with the decoding and list option. In Sec.~\ref{sec:sw}, we show that the channel coding results carry over to the Slepian-Wolf problem~\cite{sw73}. All the proofs are detailed in Sec.~\ref{sec:proofs}.
\section{Problem Setting and Main Definitions} \label{sec:prob_set}
Let $W$ be a random transformation (channel) from a discrete input alphabet $\calX$ to a discrete output alphabet $\calY$. We denote length-$n$ deterministic (resp.\ random) strings $\bx =(x_1,\ldots, x_n) \in\calX^n$  (resp.\ $\bX=(X_1, \ldots, X_n)\in\calX^n$) by lower case (resp.\ upper case) boldface. If $W^n$ satisfies $W^n(\by|\bx) = \prod_{j=1}^n W(y_j|x_j)$ for every $(\bx,\by) \in \calX^n\times\calY^n$ and the sets $\calX$ and $\calY$ are finite, $W^n$ is said to be a DMC. We focus on DMCs in this paper but  extensions to other channels such as the AWGN channel are straightforward.  For a sequence $\bx$, its {\em type} is the empirical distribution $P_{\bx}(x) = \frac{1}{n}\sum_{i=1}^n \bone\{x_i = x\}$.  For a finite alphabet $\calX$, let $\calP(\calX)$ and $\calP_n(\calX)$ be the set of probability mass functions and $n$-types~\cite{Csi97} (types with denominator at most $n$) respectively.  For two sequences $(\bx,\by)\in\calX^n\times\calY^n$, we say that $U$ is a {\em conditional type} of $\by$ given $\bx$ if $P_{\bx}(x)U(y|x)  = P_{\bx,\by}(x,y)$ for all $(x,y)\in\calX\times\calY$. ($U$ is not unique if $P_{\bx}(x)=0$ for some $x\in\calX$.) The {\em $U$-shell} of a sequence $\bx \in\calX^n$, denoted as $\calT_U(\bx)\subset\calY^n$ is the set of all $\by$ such that the joint  type of $(\bx,\by)$ is  $P_{\bx}\times U$.  The set of all conditional types $U$ for which $\calT_U(\bx)$ is non-empty for some $\bx$ with type $P$ is denoted as $\calU_n(\calY;P)$.  All logs are to the base $2$ with the understanding that $\exp(t)=2^t$. 

For information-theoretic quantities, we will  mostly follow the notation in Csisz\'ar and K\"orner~\cite{Csi97}.  We denote the {\em  information capacity} of the DMC $W:\calX\to\calY$ as  $C=C(W) := \max_{P \in \calP(\calX)}I(P,W)$. We let $\Pi =\Pi(W):=\{P\in\calP(\calX): I(P,W)=C\}$  be the set of  {\em capacity-achieving input distributions}. If $(X,Y)$ has joint distribution  $P(x)  W(y|x)$, define  $PW(y):=\sum_x P(x)W(y|x)$ to be the induced output distribution and
\begin{equation}
V(P,W):= \rmE_{X}\bigg[\var\Big( \log\frac{W(Y|X)}{PW(Y) } \,  \Big| \,X \Big) \bigg]
\end{equation}
 to be the  {\em conditional information variance}.  The {\em $\epsilon$-dispersion} of the DMC $W$ \cite{Strassen,Hayashi09,PPV10} is defined as 
\begin{align}
V_\epsilon := \left\{  \begin{array}{cc}
 V_{\min} := \min_{P\in\Pi} V(P,W) & \epsilon<1/2 \\
  V_{\max} := \max_{P\in\Pi} V(P,W) & \epsilon\ge 1/2  
\end{array}  \right. . 
\end{align}
We will assume throughout that the DMC $W$ satisfies $V_\epsilon>0$.  For integers $l\le m$, we denote $[l:m]:=\{l, l+1, \ldots,   m\}$ and $[m]:=[1:m]$.  Let $\Phi(x) := \int_{-\infty}^x (2\pi)^{-1/2}\rme^{-t^2/2}\, \rmd t$ be the cdf of a standard Gaussian and $\Phi^{-1}(\cdot)$   its inverse. We now   define   erasure codes.

\begin{definition}
An {\em $M$-erasure code} for $W:\calX\to\calY$ is a pair of mappings $(f,\varphi)$ such that $f:[M]\to\calX$ and $\varphi:\calY\to [ 0:M]$. The disjoint {\em decoding regions} are denoted as $\calD_m:=\varphi^{-1}(m)$; the {\em erasure region} is denoted as $\calD_0 :=\varphi^{-1}(0)$; and the {\em conditional undetected},  {\em erasure} and {\em total error probabilities} are defined as 
\begin{align}
\eeu(m) &:= \sum_{\tilm \in [M]\setminus\{m\}}W( \calD_{\tilm} |f(m)) \\
\eee(m) &:=  W(\calD_{0} |f(m)) \\
\eet(m) &:= \sum_{\tilm \in [0:M]\setminus\{m\}}W( \calD_{\tilm} |f(m))
\end{align}
\end{definition} 
Note that $\eeu(m)+\eee(m)=\eet(m)$. 
Typically, the code is designed so that $\eeu(m)\ll \eee(m)$ as the cost of making an undetected error is much higher than that of declaring an erasure. 
\begin{definition} \label{def:erasure_code}
An $(M,\eu, \et)_{\rma, \rmb}${\em -erasure code} for $W$ is  an $M$-erasure code for the same channel where 
\begin{enumerate}
\item If $(\rma, \rmb) = (\max,\max)$,
\begin{equation}
\max_{m \in [M]}\eeu(m)\le\eu,\quad \max_{m \in [M]}\eet(m)\le\et.
\end{equation}
\item If   $(\rma, \rmb) = (\max,\mathrm{ave})$ 
\begin{equation}
\max_{m \in [M]}\eeu(m)\le\eu,\quad  \frac{1}{M}\sum_{m \in [M]}\eet(m)\le\et.
\end{equation}

\item If $(\rma, \rmb) = (\mathrm{ave},\max)$,
\begin{equation}
\frac{1}{M}\sum_{m \in [M]}\eeu(m)\le\eu,\quad \max_{m \in [M]}\eet(m)\le\et.
\end{equation}

\item If $(\rma, \rmb) = (\mathrm{ave},\mathrm{ave})$,
\begin{equation}
\frac{1}{M}\sum_{m \in [M]}\eeu(m)\le\eu,\quad \frac{1}{M}\sum_{m \in [M]}\eet(m)\le\et.
\end{equation}
\end{enumerate}
\end{definition}
In Definition~\ref{def:erasure_code}, we consider  erasure codes with constraints on  the {\em undetected} and {\em total} error probabilities similar to~\cite{merhav08,Forney68,Moulin09}. An alternate formulation would be to consider  $(M,\eu, \ee)_{\rma, \rmb}$-erasure codes where $\ee$ is the {\em erasure} probability. We find the former formulation more traditional and the analysis is also somewhat easier.

\begin{definition}
A number $r\in\bbR$ is an  {\em $(\eu,\et)_{\rma, \rmb}$-achievable erasure second-order coding rate} for the DMC $W^n$ with capacity $C$ if there exists a sequence of $(M_n,\eun, \etn)_{\rma, \rmb}${\em -erasure codes}      such that 
\begin{align}
\liminf_{n\to\infty}\frac{1}{\sqrt{n}}(\log M_n -nC)& \ge r ,  \label{eqn:Mcond}\\
\limsup_{n\to\infty}\eun &\le \eu,  \quad\mbox{and}    \\
\limsup_{n\to\infty}\etn  & \le \et. \label{eqn:etn}
\end{align}
The {\em $(\eu,\et)_{\rma, \rmb}$-erasure second-order capacity}  $r^*_{\mathrm{era}, \rma,\rmb}(\eu,\et)$ is the supremum of all  $(\eu,\et)_{\rma, \rmb}$-achievable erasure second-order coding rates.
\end{definition}
We now turn our attention to codes which allow their decoders to output a {\em list} of messages.  Let $\binom{[M]}{j}$ be the set of subsets of $[M]$ of size $j$. Furthermore, we use the notation $\binom{[M]}{\le L} :=\cup_{0\le j\le L}  \binom{[M]}{j}$ to denote the set of subsets of $[M]$ of size not exceeding $L$. 
\begin{definition}
An {\em $(M,L)$-list code}  for $W:\calX\to\calY$ is a pair of mappings $(f,\varphi)$ such that $f:[M]\to\calX$ and $\varphi:\calY\to\binom{[M]}{\le L}$. The  (not-necessarily disjoint) {\em decoding regions} are denoted as $\calD_m:= \{y\in\calY:m\in\varphi(y)\}$   and the {\em conditional error probability} is defined as 
\begin{align}
\lambda(m)  := W (\calY\setminus\calD_m |f(m)) .
\end{align}
\end{definition}

\begin{definition} \label{def:list_codes}
An {\em $(M,L,\epsilon)_{\rma}$-list code}  for $W$ is an $(M,L)$-list code for the same channel where if $\rma=\max$, 
\begin{equation}
\max_{m \in [M]}\lambda(m) \le \epsilon,
\end{equation}
or if $\rma=\mathrm{ave}$, 
\begin{equation}
\frac{1}{M}\sum_{m \in [M]}\lambda(m) \le \epsilon.
\end{equation}
\end{definition}

\begin{definition} \label{eqn:list_second}
A number $r\in\bbR$ is an  {\em $(l,\epsilon)_{\rma}$-achievable list second-order coding rate} for the DMC $W^n$ with capacity $C$ if there exists a sequence of $(M_n,L_n, \epsilon_n)_{\rma}${\em -list codes}      such that  in addition to \eqref{eqn:Mcond}, the following hold
\begin{align}
\limsup_{n\to\infty}\frac{1}{\sqrt{n}}\log L_n &\le l,  \quad\mbox{and}\label{eqn:list_size} \\
 \limsup_{n\to\infty}\epsilon_n  &\le \epsilon. \label{eqn:error_no_rootn}
\end{align}
The {\em $(l,\epsilon)_{\rma}$-list second-order capacity} $r^*_{\mathrm{list}, \rma}(l,\epsilon)$ is the supremum of all  $(l,\epsilon)_{\rma}$-achievable list second-order coding rates.
\end{definition}

According to~\eqref{eqn:list_size}, we stipulate that the list size grows as $\exp(  \sqrt{n} \, l )$. This differs from previous works on list decoding in which the list size is either constant~\cite{gallagerIT, merhav13} or exponential~\cite{Csi97, tel94, Moulin09, sgb, merhav08, merhav13, Blinovsky, merhav13a}. This scaling affects the second-order (dispersion) term. To understand how the list size may affect the higher-order term, consider the following definition.   

\begin{definition} \label{def:list_third}
A number $s\in\bbR$ is an {\em $(\alpha,\epsilon)$-achievable list third-order coding rate} for the DMC $W^n$ with capacity $C$ and positive $\epsilon$-dispersion $V_\epsilon$ if there   exists a sequence of $(M_n,L_n, \epsilon_n)_{\mathrm{ave}}${\em -list codes}   such that
\begin{align}
\liminf_{n\to\infty}\frac{1}{\log n}  \big(\log M_n-nC-\sqrt{nV_{\epsilon_n}}\Phi^{-1}(\epsilon_n) \big)  &\ge s,   \label{eqn:divide_log}\\
\limsup_{n\to\infty} \frac{\log L_n}{\log n} &\le \alpha ,  \quad\mbox{and} \label{eqn:list_size_poly}\\
 \limsup_{n\to\infty} \sqrt{n}(\epsilon_n -\epsilon) & <\infty. \label{eqn:error_rootn}
\end{align}  
The {\em $(\alpha,\epsilon)$-list third-order capacity} $s^*_{\mathrm{list}}(\alpha,\epsilon)$ is the supremum of all $(\alpha,\epsilon)$-achievable list third-order coding rates.
\end{definition}

Inequality  \eqref{eqn:list_size_poly} implies that the size of the list grows polynomially and in particular it  scales as $O(n^\alpha)$.  This scaling affects the third-order (logarithmic) term studied by a number of authors~\cite{TomTan12, mou13b, TanTom13, altug13} in the context of ordinary channel coding (without list decoding). By \eqref{eqn:divide_log},  if $s\in\bbR$ is  $(\alpha,\epsilon)$-achievable,  there exists a sequence of codes of sizes $M_n$  satisfying
\begin{equation}
\log M_n\ge nC + \sqrt{nV_{\epsilon_n}}\Phi^{-1}(\epsilon_n) + s\log n + o(\log n), \label{eqn:ologn}
\end{equation}
and having list sizes $L_n = O(n^\alpha)$ and average error probabilities $\epsilon_n$ not exceeding $\epsilon+ O(n^{-1/2})$. 
The more stringent condition on the sequence of error probabilities $\{\epsilon_n\}_{n\ge 1}$ in \eqref{eqn:error_rootn} (relative to  \eqref{eqn:error_no_rootn}) is because $\epsilon_n$ appears as the argument in the $\Phi^{-1}(\cdot)$ function, which forms part of the coefficient of the $\sqrt{n}$ second-order term in the asymptotic expansion of $\log M_n$. If the weaker condition \eqref{eqn:error_no_rootn} were in place, the approximation error between $\epsilon_n$ and the target $\epsilon$ which  is of the order $o(\sqrt{n})$ would affect the third-order term, which is the object of study here. The stronger condition in \eqref{eqn:error_rootn} ensures that the third-order (logarithmic) term is unaffected by the approximation error between $\epsilon_n$ and $\epsilon$ which is now of the order $O(1)$. 

\section{Main Results for Channel Coding} \label{sec:main_res}
In this section, we summarize the main results of this paper concerning channel coding with an erasure or list option. For simplicity, we assume that the DMC $W$ satisfies $V_{\epsilon}>0$  though our results can be extended in a straightforward manner to the case where $V_{\epsilon}=0$. 
\subsection{Decoding with Erasure Option}
\begin{theorem} \label{thm:erasure}
For any $0  \le  \eu <  \et< 1$, 
\begin{equation}
r^*_{\mathrm{era}, \rma,\rmb}(\eu,\et)=\sqrt{V_{\et}}\Phi^{-1}(\et), \label{eqn:thm_erasure}
\end{equation}
where $(\rma,\rmb)$  can be any element in  $\{\max,\mathrm{ave}\}^2$. 
\end{theorem}
The proof of Theorem~\ref{thm:erasure} can be found in Section~\ref{sec:prf_erasure}. 

A few comments are in order: First,  Theorem~\ref{thm:erasure} implies that if $M^*(W^n;\eu,\et)$ is the maximum number of codewords that can be transmitted over $W^n$ with undetected and total error $\eu$ and $\et$ respectively, then 
\begin{equation}
\log M^*(W^n;\eu,\et)=nC + \sqrt{nV_{\et}}\Phi^{-1}(\et) + o(\sqrt{n}). \label{eqn:backoff_era}
\end{equation}
We see that the backoff from the capacity at blocklength $n$ is   approximately  $-\sqrt{{V_{\et}}/{n}}\, \Phi^{-1}(\et)$ independent of $\eu$. (This backoff is positive for $\et<1/2$.) 
Observe that the second-order term does not depend  on $\eu$, the undetected error probability. Only the total error probability comes into play in the asymptotic characterization of $\log M^*(W^n;\eu,\et)$.   In fact, in the proof, we first argue that it suffices to show that  $\sqrt{V_{\ee}}\Phi^{-1}(\ee)$ is an achievable $(0, \ee)_{\mathrm{ave}, \mathrm{ave}}$-erasure second-order coding rate for $W^n$, i.e., the undetected error probability is asymptotically $0$.  Clearly, any achievable  $(0, \et)_{\mathrm{ave}, \mathrm{ave}}$-erasure second-order coding rate is also  an achievable $(\eu, \et)_{\mathrm{ave}, \mathrm{ave}}$-erasure second-order coding rate  for any $\eu\in [0,\et)$. 

Second, in the direct part of the proof of Theorem \ref{thm:erasure}, we  use  threshold decoding, i.e.\ declare that message  $m$ is sent if the empirical mutual information     is higher than a threshold. If no message's empirical mutual information  exceeds the threshold, then an erasure is declared. This simple rule, though {\em universal}, is not the optimal one (in terms of minimizing the total error while holding the undetected error fixed). The optimal rule was derived using a generalized version of the Neyman-Pearson lemma by Forney~\cite[Thm.~1]{Forney68} and it is stated as
\begin{equation}
\calD_m^*  :=  \bigg\{\by: W^n(\by|f(m))\ge\psi \sum_{\tilm \in [M]\setminus\{m\} } W^n(\by|f(\tilm)) \bigg\},\label{eqn:opt_rule}
\end{equation}
for some threshold $\psi>0$. However, this rule appears to be difficult for second-order analysis and is more amenable to error exponent analysis~\cite{Moulin09, merhav13}. Because the  likelihood of the second most likely codeword   is usually much higher than the rest (excluding the first), Forney  also suggested  the  simpler but, in general, suboptimal  rule~\cite[Eq.~(11a)]{Forney68}
\begin{equation}
\calD_m'  :=  \left\{\by :  W^n(\by|f(m)) \ge \psi  \max_{\tilm \in [M]\setminus\{m\} }  W^n(\by|f(\tilm)) \right\}. \label{eqn:simpler_rule}
\end{equation}
We analyzed this rule  in the asymptotic setting (i.e., as $n$ tends to infinity) in the same way as one analyzes the random coding union (RCU) bound~\cite[Thm.~16]{PPV10}  \cite[Sec.~7]{mou13b} for ordinary channel coding but the  analysis is more involved than threshold decoding which suffices for proving the second-order achievability of~\eqref{eqn:thm_erasure}. What was somewhat surprising to the authors is that the optimal decoding scheme in \eqref{eqn:opt_rule} (a generalization of the Neyman-Pearson lemma to arbitrary non-negative measures) and the suboptimal scheme in \eqref{eqn:simpler_rule} are somewhat difficult to analyze,  but the simpler empirical mutual information thresholding rule can be shown to be second-order optimal. This is in contrast to error exponent analysis of erasure decoding where the rules \eqref{eqn:opt_rule} and \eqref{eqn:simpler_rule} and their variants are ubiquitously analyzed in the literature, e.g.,~\cite{Csi97, tel94, Moulin09,  merhav08}.


Third, the converse is based on Strassen's idea~\cite[Eq.~(4.18)]{Strassen}, establishing a clear link between point-to-point channel coding  and  binary hypothesis testing. Also see Kemperman's general converse bound~\cite[Lem.~3.1]{Kemperman} and for a more modern treatment, the various forms of the {\em meta-converse} in~\cite[Sec.~III-E]{PPV10}. The hypothesis testing converse technique only depends on the total error probability, explaining the presence of $\et$ and not $\eu$ in~\eqref{eqn:thm_erasure}.

Finally, we remark that Theorem~\ref{thm:erasure} (as well as Theorem~\ref{thm:list} to follow bar the statement in \eqref{eqn:s_third2}) carries over verbatim for the AWGN channel where  the capacity and dispersion are $C=\frac{1}{2}\log (1+S)$ and  $V =\log^2\rme\cdot\frac{S(S+2)}{2(S+1)^2}$   respectively and $S$ is the signal-to-noise ratio (SNR) of the AWGN channel.

\subsection{The Expected Rate and An Example} \label{sec:exp_rate}
We now compare the expected rate achieved using decoding with the erasure option to ordinary channel coding. Define
\begin{equation}
\ee:=\et-\eu
\end{equation}
as the {\em erasure probability} and note from the assumption of Theorem \ref{thm:erasure} that $\ee>0$. Now consider sending $b \in\bbN$ {\em independent} blocks of information each of length $n\in\bbN$. Because  transmission succeeds (no erasure declared) with probability $1-\ee$,  the total number of bits we can transmit in each block is well approximated by the random variable 
\begin{equation}
R_{\rme}^{(n)} :=\left\{ \begin{array}{cl}
C + \sqrt{ V_{\et} /{n}  }  \, \Phi^{-1}(\et) & \mathrm{w.p.} \,\, \,\,1-\ee \\
0 &  \mathrm{w.p.} \,\,\,\, \ee 
\end{array}  \right. \label{eqn:Ren}
\end{equation}
This random variable has expectation 
\begin{equation}
\rmE\big[ R_{\rme}^{(n)} \big] := (1-\ee)\bigg(C + \sqrt{\frac{V_{\et}}{n}}\Phi^{-1}(\et)\bigg). \label{eqn:def_er}
\end{equation}
Fix $\delta\in (0,\min\{\ee,1-\ee\})$. By Hoeffding's inequality, the total number of bits we can transmit over the $b$ blocks  is in  the interval
\begin{equation}
\left[ (1-\ee-\delta)\cdot b\cdot   \big(nC + \sqrt{n{V_{\et}}} \Phi^{-1}(\et)\big) , (1-\ee +\delta)\cdot b\cdot  \big(nC + \sqrt{n{V_{\et}}} \Phi^{-1}(\et)\big)   \right]\label{eqn:interval}
\end{equation}
with  probability  exceeding $1-2\,  \rme^{-b\delta^2 }$. 
This reduction in rate in \eqref{eqn:def_er}--\eqref{eqn:interval} by the factor of $1-\ee$ in the so-called {\em single-codeword repetition} scheme    was first  observed by Forney~\cite[Eq.~(49)]{Forney68}. Essentially, one may use an {\em automatic repeat request (ARQ)} scheme\footnote{Forney~\cite{Forney68} calls this class of retransmission schemes {\em decision feedback} schemes in his paper, but nowadays the term ARQ is more common.} to resend the entire block of information if there is an erasure. 

\begin{figure} 
\centering
\includegraphics[width = 0.95\columnwidth]{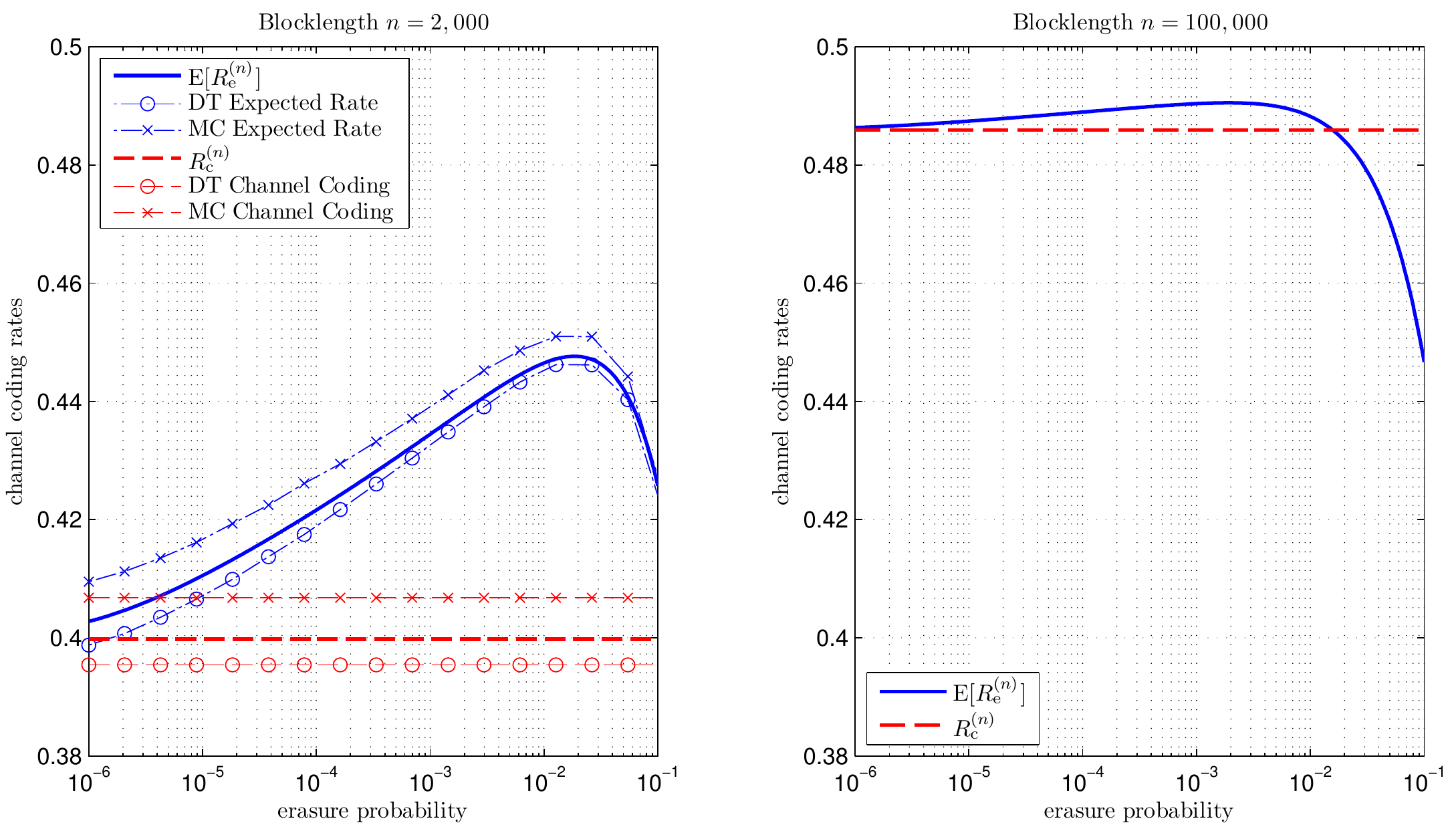} 
\caption{Comparison of non-asymptotic rates and Gaussian approximationss for channel coding with and without  the erasure option. Observe that  $\rmE [ R_{\rme}^{(n)} ]$ can be larger $ R_{\rmc}^{(n)} $ for finite $n$. On the left plot, DT Expected Rate and MC Expected Rate respectively stand for the dependence-testing (DT)~\cite[Thm.~34]{PPV10} and meta-converse (MC)~\cite[Thm.~35]{PPV10} finite blocklength bounds for the expected rate with the given parameters $(\ee,\eu,n,q)$. DT Channel  Coding and MC Channel  Coding  respectively stand for the DT and MC finite blocklength bounds for channel coding with ordinary decoding with the given parameters $(\eu,n,q)$.  The finite blocklength bounds are difficult to compute   numerically  for large blocklengths and small error probabilities and so they are not shown  on the right plot.}
\label{fig:eras}
\end{figure}

For ordinary channel coding with error probability  $\eu$,  we can send approximately 
\begin{equation}
b\cdot \big(nC + \sqrt{nV_{\eu}}\Phi^{-1}(\eu) \big)
\end{equation}
bits over the $b$ independent blocks and so, dividing by $nb$,   the  non-asymptotic  channel coding rate (analogue of~\eqref{eqn:def_er}) can be approximated~\cite{PPV10}~by 
\begin{equation}
R_{\rmc}^{(n)} := C + \sqrt{\frac{V_{\eu}}{n}}\Phi^{-1}(\eu). \label{eqn:Rc}
\end{equation}
In the analysis in \eqref{eqn:Ren}--\eqref{eqn:Rc}, we have assumed that the Gaussian approximation is sufficiently accurate.  It was numerically shown in~\cite{PPV10} that   the Gaussian approximation is accurate for some channels (such as the binary symmetric channel, binary erasure channel and additive white Gaussian noise channel) and moderate blocklengths (of order $\approx 100$) and error probabilities (or order $\approx 10^{-6}$). Also see Fig.~\ref{fig:eras} for a precise quantification of the accuracy of the Gaussian approximation in our setting. 

Clearly if $0\le \eu<\ee \le  \et<1$ are constants,  
\begin{equation}
C=\lim_{n\to\infty} R_{\rmc}^{(n)} > \lim_{n\to\infty} \rmE\big[ R_{\rme}^{(n)} \big]=(1-\ee)C, \label{eqn:n_large}
\end{equation}
so there is no advantage in allowing for erasures asymptotically. However, in finite blocklength (by this we mean the per-block blocklength $n$) regime, for ``moderate'' $\ee$, we may have 
\begin{equation}
 R_{\rmc}^{(n)} < \rmE\big[ R_{\rme}^{(n)} \big]
\end{equation}
so erasure decoding may be advantageous {\em in expectation}.     We illustrate the difference between $\rmE\big[ R_{\rme}^{(n)} \big]$  and $R_{\rmc}^{(n)}$ with a concrete example.
 
{\bf Example}: In   Fig.~\ref{fig:eras}, we consider  a binary symmetric channel (BSC) with crossover probability $q = 0.11$ so $C = 0.5$ bits/channel use and $V = 0.891$ bits$^2$/channel use.  (For the BSC, $V_\epsilon$ does not depend on $\epsilon$.) We keep the undetected error probability at $\eu=10^{-6}$ and vary the erasure error probability  $\ee$ in the interval $[10^{-6},  10^{-1}]$. We chose two blocklengths $n\in \{2\times 10^3, 1\times 10^5\}$.  We observe that for blocklength $n = 2\times 10^3$ and a moderate erasure probability of  $\ee\approx 10^{-2}$, the gain of  coding with the erasure option over ordinary channel coding is rather pronounced. This can be seen by comparing either the Gaussian approximations or the finite blocklength bounds. This gain is reduced if (i) $\ee$ is increased because we retransmit the whole block more often on average via the use of decision feedback or (ii) $n$ becomes large so the second-order $\sqrt{n}$ term becomes less significant (cf.~\eqref{eqn:n_large}). We also note from the left plot that the Gaussian approximation is reasonably accurate when compared to the  dependence-testing~\cite[Thm.~34]{PPV10} and meta-converse~\cite[Thm.~35]{PPV10} finite blocklength bounds under the current settings. The DT bound is especially close to the Gaussian approximation when the performance of coding with erasures peaks.

Finally, we remark that the advantage of coding with the erasures option over ordinary channel coding was also shown from the error exponents perspective by Forney in \cite{Forney68}. More precisely, Forney~\cite[Eq.~(55)]{Forney68} showed that the feedback exponent has slope $-1$ for rates near and below capacity, thus improving on the ordinary exponent which has slope $0$ in the same region.

\subsection{List Decoding}
\begin{theorem} \label{thm:list}
For any  $0<\epsilon<1$,    we have the second-order result
\begin{equation}
r^*_{\mathrm{list}, \rma}(l,\epsilon)=l+\sqrt{V_{\epsilon}}\Phi^{-1}(\epsilon). \label{eqn:r_second}
\end{equation}
where $\rma\in \{\max,\mathrm{ave}\}$.  Furthermore, if $\alpha \ge 0$ and $0<\epsilon<1$, we  also have the third-order result
\begin{equation}
\alpha\le s^*_{\mathrm{list}}(\alpha,\epsilon) \le\frac{1}{2}+\alpha . \label{eqn:s_third}
\end{equation}
 If the DMC $W$ is symmetric in the Gallager sense\footnote{A DMC is {\em symmetric in the Gallager sense}~\cite[Sec.~4.5, pp.~94]{gallagerIT} if the set of channel outputs $\calY$ can be partitioned into subsets such that within each subset, the matrix of transition probabilities satisfies the following: every row (resp. column) is a permutation of every other row (resp. column).} and singular,\footnote{A DMC $W$ is {\em singular} if for all $(x,y,\barx) \in \calX\times\calY\times\calX$, with $W(y|x)W(y|\barx)>0$ it is true that $W(y|x)=W(y|\barx)$.  }  then we can make the stronger statement 
\begin{equation}
s^*_{\mathrm{list}}(\alpha,\epsilon)=\alpha. \label{eqn:s_third2}
\end{equation}
\end{theorem}
The proof of Theorem~\ref{thm:list} which is partly based on Proposition \ref{prop:list} below can be found in Section~\ref{sec:prf_list}. 

Theorem~\ref{thm:list} shows that if we allow the list to grow as $\exp(\sqrt{n}\, l)$, then the second-order capacity is increased by $l$. This is   concurs with the intuition we obtain from the analysis of error exponents for list decoding by Shannon-Gallager-Berlekamp~\cite{sgb}. See also the exercises in Gallager~\cite[Ex.~5.20]{gallagerIT} and Csisz\'ar-K\"orner~\cite[Ex.~10.28]{Csi97} which are concerned solely with first-order (capacity) analysis. 

For the third-order result in \eqref{eqn:s_third}, we observe that there is a gap. This is, in part, due to the fact that we use threshold decoding to decide which messages belong to the list. This decoding strategy appears to be suboptimal in the third-order sense for  DMCs \cite[Sec.~7]{mou13b} \cite[Thm.~53]{Pol10} and AWGN channels~\cite{TanTom13}. It also appears that one needs to use a version of maximum-likelihood (ML) decoding as in \eqref{eqn:simpler_rule} and analyze an analogue of the RCU bound~\cite[Thm.~16]{PPV10} carefully   to obtain the additional $\frac{1}{2}$ for the direct part. Whether  this can be done for channel coding with list decoding to obtain a tight third-order result general DMCs   is an open question.  Nonetheless for singular and symmetric channels considered by Altu\u{g} and Wagner~\cite{altug13}, such as the BEC, the converse (upper bound)  can be tightened and this results in the conclusive result  in~\eqref{eqn:s_third2}.

A by-product of the proof of Theorem~\ref{thm:list} is the following  non-asymptotic converse bound for list-decoding which may be of independent interest. Kostina-Verd\'u~\cite[Thm.~4]{kost13}   developed and used a version of this bound for the purposes of  fixed error asymptotics of joint source-channel coding, just as Csisz\'ar~\cite{Csi80} also used a list decoder in his study of the error exponents for joint source-channel coding. 
\begin{proposition}\label{prop:list}
Let $\beta_\alpha(P,Q)$ be the best (smallest) type-II error in a (deterministic) hypothesis test between $P$ and $Q$ subject to the type-I error  being no larger than $1-\alpha$. Every $(M,L,\epsilon)_{\mathrm{ave}}$-list code for $W:\calX\to\calY$ satisfies
\begin{align}
 \frac{M}{L}\le\inf_{Q \in\calP(\calY)}\sup_{P \in\calP(\calX)}  \frac{1}{\beta_{1-\epsilon}(P\times W ,  P\times Q)}. \label{eqn:meta_list}
\end{align}
\end{proposition}
This bound immediately reduces to the so-called {\em meta-converse} in \cite[Sec.~III-E]{PPV10} by setting $L=1$. We will see that the {\em ratio} $M/L$ plays a  critical role in both the converse and direct parts. This is also evident in existing works such as Shannon-Gallager-Berlekamp~\cite[Sec.~IV]{sgb} but the non-asymptotic bound in \eqref{eqn:meta_list} appears to be novel.


\section{An Extension to Slepian-Wolf Coding} \label{sec:sw}
In this section, we show that the techniques developed are also applicable to lossless source coding with decoder side information, i.e., the Slepian-Wolf problem~\cite{sw73}. Let $P_{XY} \in \calP(\calX\times\calY)$ be a correlated source where the alphabets $\calX$ and $\calY$ are finite.  The assumption that  $\calX$ and $\calY$ are finite sets can be dispensed at the cost of non-universality in the coding scheme~\cite{Nom13}.
\begin{definition}
An {\em $M$-code} for the correlated source $P_{XY}$ is defined by two functions $f: \calX\to [M]$ and $\varphi : [M]\times \calY\to \calX\cup\{ \emptyset\}$ where $\emptyset$ represents the erasure symbol.
\end{definition}

\begin{definition}
An {\em $(M,\eu,\et)$-code} for the correlated source $P_{XY}$ is an $M$-code satisfying 
\begin{align}
\sum_{x,y}P_{XY}(x,y) \bone\big\{ \varphi (f(x),y)  \in \calX\setminus\{x\}\big\} & \le \eu,\quad\mbox{and}\\
\sum_{x,y}P_{XY}(x,y) \bone\big\{ \varphi (f(x),y) \ne  x\big\} & \le \et.
\end{align}
The parameters $\eu$ and $\et$ are known as the {\em undetected} and {\em total error probabilities} respectively.
\end{definition}

We consider discrete, stationary and memoryless sources $P_{\bX\bY}(\bx,\by) = \prod_{j=1}^n P_{XY}(x_j, y_j)$ in which the source alphabet is $\calX^n$ and the side-information alphabet is $\calY^n$. 

\begin{definition}
A number $r \in\bbR$ is said to be an {\em $(\eu,\et)$-achievable second-order coding rate} for the correlated source $\{P_{X^n Y^n}\}_{n\ge 1}$ if there exists a sequence of $(M_n, \eun, \etn)$-codes such that 
\begin{align}
\limsup_{n\to\infty} \frac{1}{\sqrt{n}} (\log M_n-nH(X|Y) ) &\le r \\
\limsup_{n\to\infty} \eun &\le\eu\\
\limsup_{n\to\infty} \etn &\le\et.
\end{align}
The infimum of all $(\eu,\et)$-achievable second-order coding rates is the {\em optimum second-order coding rate $r_{\mathrm{era}}^*(\eu,\et)$}. 
\end{definition}

Hence, we are allowing the erasure option for the Slepian-Wolf problem but we restrict the undetected and total errors to be at most $\eu$ and $\et$ respectively. The second-order asymptotics for the Slepian-Wolf problem with two encoders  without erasures was studied by Tan and Kosut in~\cite{TK14}.  

Define $V(X|Y) := \var(-\log P_{X|Y}(X|Y))$ to be  the {\em conditional varentropy}~\cite{verdu14} of the stationary, memoryless source $P_{XY}$.    The following result here parallels Theorem~\ref{thm:erasure}   pertaining to channel coding with the erasure option.

\begin{theorem} \label{thm:sw}
For any  $0  \le  \eu <  \et< 1$, 
\begin{equation}
r_{\mathrm{era}}^*(\eu,\et)=\sqrt{V(X|Y)} \Phi^{-1}(1-\et).
\end{equation}
\end{theorem}
The proof of Theorem~\ref{thm:sw} can be found in Section~\ref{sec:prf_sw}. The direct part uses random binning~\cite{cover75} and thresholding of the empirical conditional entropy~\cite{TK14} and the converse uses a  non-asymptotic information spectrum  converse bound by Miyake and Kanaya~\cite{miyake}. Also see \cite[Thm.~7.2.2]{Han10}.

Again, we observe that the optimum second-order coding rate does not depend on the undetected error probability $\eu$ as long as it is  smaller than the total error probability $\et$. Hence, the observation we made in Fig.~\ref{fig:eras}--namely that at finite blocklengths the expected performance with the erasure option can exceed that without the erasure option--also applies in the Slepian-Wolf setting.

\section{Proofs} \label{sec:proofs}
In this section, we provide the proofs of  the theorems in the paper. 
\subsection{Decoding with Erasure Option: Proof of Theorem~\ref{thm:erasure}} \label{sec:prf_erasure}
\subsubsection{Converse Part}  \label{sec:conv_prf_erasure}  Recall that $\beta_\alpha (P,Q)$ is  the best (smallest) type-II error in a   hypothesis test (without randomization) of $P$ versus $Q$ subject to the condition that the type-I error is no larger than $1-\alpha$. This function was studied extensively in \cite{Pol13}.  Let us fix any $(M_n,\eun, \etn)_{\max,\max}$-erasure code for $W^n$. This means that
\begin{equation}
\min_{m\in [M_n]} W^n (\calD_m | f(m) )  \ge 1-\etn. \label{eqn:total_prob}
\end{equation}
Note that only the total error comes into play in  \eqref{eqn:total_prob} and thus the second-order capacity in \eqref{eqn:thm_erasure} only depends on $\et$. In essence, an $(M_n,\eun, \etn)_{\max,\max}$-erasure code for $W^n$ is an $(M_n,  \etn)_{\max}$-channel code for $W^n$ (a channel code for $W^n$ with $M_n$ codewords and maximum error probability at most $\etn$) so any converse for usual channel coding also applies here with the error probability for usual channel coding being the total error in erasure decoding.  We describe some details of the converse proof to make the paper self-contained.

We now assume that the code is constant composition, i.e.\ all codewords are of the same type $P$. This only leads to a $O(\log n)$ penalty in $\log M_n$ which does not affect the second-order term.  Now, for any permutation invariant output distribution $Q \in\calP(\calY^n)$ (this means that $Q(y_1, \ldots, y_n)=Q(y_{\pi(1)},\ldots, y_{\pi(n)})$ for every permutation $\pi :[n]\to [n]$), we have 
\begin{align}
1 &\ge \sum_{m=1}^{M_n} Q(\calD_m) \label{eqn:disjoint_sets}  \\
&\ge\sum_{m=1}^{M_n} \min_{\substack{\calD\subset\calY^n: \\ W^n(\calD|f(m))\ge 1-\etn}} Q(\calD)  \label{eqn:satisfies_bd}\\
  & =  \sum_{m=1}^{M_n} \beta_{1-\etn} (W^n(\cdot | f(m)) , Q)   \label{eqn:use_def_beta} \\
&  =  {M_n}\beta_{1-\etn} (W^n(\cdot | \bx) , Q) ,  \label{eqn:one_seq}
\end{align}
where  \eqref{eqn:disjoint_sets} follows because $\calD_m , m \in [M_n]$ are disjoint; \eqref{eqn:satisfies_bd} follows from \eqref{eqn:total_prob}; \eqref{eqn:use_def_beta} uses the definition of $\beta_\alpha$;  and  finally~\eqref{eqn:one_seq} follows from the fact that  $\beta_{1-\etn} (W^n(\cdot | f(m)) , Q)$ does not depend on $m$ for permutation invariant $Q$ (which is what we choose $Q$ to be) and constant composition codes~\cite{Strassen}. We also used $\bx$  to denote any element in the type class $\calT_P$. Choose $Q$ to be the product distribution $(PW)^n$. Since $\etn$ satisfies  \eqref{eqn:etn}, for every $\eta \in (0, 1-\et)$, there exists sufficiently large $n$ such that $\etn\le\et+\eta$. Hence, from~\eqref{eqn:one_seq},
\begin{equation}
M_n\le \beta_{1-\et-\eta}^{-1} (W^n(\cdot|\bx) , Q). \label{eqn:boundM}
\end{equation}
Now, we use \cite[Sec.~2]{Strassen} to assert that  for all $\epsilon\in (0,1)$, 
\begin{align}
-\log     \beta_{1-\epsilon}(W^n(\cdot|\bx),Q)  
  = n I(P,W)+\sqrt{nV(P,W)}\Phi^{-1}(\epsilon)+O(\log n  ).  \label{eqn:boundbeta}
\end{align}
Hence, by putting \eqref{eqn:boundM} and \eqref{eqn:boundbeta} together, we obtain
\begin{align}
 \log M_n \le \max_{P \in \calP(\calX)}n I(P,W)&+\sqrt{nV(P,W)}\Phi^{-1}(\et+\eta) +O(\log n  )
\end{align} 
By using the usual continuity arguments (e.g.~\cite[Lem.~7]{TomTan12}), 
\begin{equation}
\log M_n \le n C + \sqrt{nV_{\et+\eta}}\Phi^{-1}(\et+\eta)+O(\log n).
\end{equation}
Thus, by letting $n\to\infty$, 
\begin{equation}
\limsup_{n\to\infty}\frac{1}{\sqrt{n}} ( \log M_n-nC )\le \sqrt{ V_{\et+\eta}}\Phi^{-1}(\et+\eta). 
\end{equation}
The converse proof is complete for the $(\max,\max)$ case by taking $\eta\downarrow 0$.   Note that even though $V_\epsilon$ is, in general, discontinuous at $\epsilon=\frac{1}{2}$ (i.e., when $V_{\min}<V_{\max}$), we may let $\eta\downarrow 0$ and use the fact that  $\Phi^{-1}(\frac{1}{2})=0$ to assert that $\lim_{\eta\to 0} \sqrt{ V_{\frac{1}{2}+\eta}}\Phi^{-1}(\frac{1}{2}+\eta)=0$ as desired.  

To get to the $(\mathrm{ave},\mathrm{ave})$ setting, first we let $M^*_{\rma,\rmb}(W, \eu, \et)$ be the maximum number of codewords $M$ in an erasure with undetected and total error probabilities $\eu,\et$ respectively under the $(\rma,\rmb)$-setting. By using an expurgation argument as in \cite[Eq.~(284)]{PPV10} it is easy to show that for all $\tau>1$, 
\begin{align}
M^*_{\mathrm{ave},\mathrm{ave}}(W^n,\eu, \et) \le    \left(\frac{1}{1-1/\tau}\right)^2  M^*_{\max,\max}(W^n,\tau\eu,\tau \et) .  \label{eqn:expur1}   
\end{align}
Setting $\tau :=1+\frac{1}{\sqrt{n}}$ proves the claim  for all the other 3 cases.
 
 
\subsubsection{Direct Part} \label{sec:era_prf_dir} It suffices to  prove that $\sqrt{V_{\ee}}\Phi^{-1}(\ee)$ is an achievable $(0, \ee)_{\mathrm{ave}, \mathrm{ave}}$-erasure second-order coding rate for $W^n$. Indeed, here the total error  $\et=\ee$. However, any achievable $(0, \et)_{\mathrm{ave}, \mathrm{ave}}$-second-order coding rate is also an achievable $(\eu, \et)_{\mathrm{ave}, \mathrm{ave}}$-second-order coding rate  for  any $\eu \in [ 0, \et)$. Furthermore, by an expurgation argument similar to \eqref{eqn:expur1}, the same statement can be proved under the $(\max,\max)$ setting.

For this proof,  we show that the second-order capacity in \eqref{eqn:thm_erasure} is also {\em universally attainable}--i.e.\ the code  does not require channel knowledge. A simpler proof based on thresholding the likelihood  can also be be used; however, channel knowledge is required. See  Sec.~\ref{sec:prf_list_direct} for an analogue of the alternative  strategy.

Fix a type $P\in\calP_n(\calX)$. Generate $M_n$ codewords uniformly at random from the type class $\calT_P$. Denote the random  codebook as $\{\bX(m):  m \in [  M_n]\} \subset\calT_P$. The number $ M_n$ is to be chosen later. Let $\gamma\ge 0$ be some threshold to be chosen later.  At the receiver, we decode  to $\hatm$ if and only if $\hatm$ is the unique message to satisfy
\begin{equation}
\hatI(\bX(\hatm)\wedge \bY)\ge\gamma
\end{equation}
where $\hatI(\bx \wedge \by)$ is the empirical mutual information of $(\bx , \by)$, i.e.\ the mutual information of the random variables $(\tilX,\tilY)$ whose distribution is the joint type $P_{\bx ,\by}$. Assume as usual that the true message $m =1$. We use the following elementary result which is     shown in the proof of the packing lemma~\cite[Lem.~10.1]{Csi97}. 

\begin{lemma} \label{lem:conc_mi}
Let $P \in \calP_n(\calX)$ be any $n$-type. Let $\bX$ and $\overline{\bX}$ be selected independently and uniformly at random from the type class $\calT_P\subset\calX^n$. Let $\bY$ be the channel output when $\bX$ is the input, i.e.\ $\bY| \{\bX =\bx\}\sim \prod_{j=1}^n W(\fndot|x_j)$. Then, for every $\gamma>0$ and every $n\in\bbN$,
\begin{align}
\Pr\big[ \hatI(\overline{\bX}\wedge \bY)>\gamma\big]\le (n+1)^{|\calX|+|\calX||\calY|} \exp(-n\gamma).
\end{align}
\end{lemma}
The undetected error probability is bounded as 
\begin{align}
\Pr[\calE_{\rmu}] &\le \Pr\left[\max_{m\in [M_n]\setminus\{1\}} \hatI(\bX(m)\wedge\bY)\ge \gamma\right] \label{eqn:maxle} \\
& \le ( M_n-1) \Pr\big[  \hatI(\bX(2)\wedge\bY)\ge \gamma\big]  \label{eqn:union_bd}\\
& \le ( M_n-1) (n+1)^{|\calX| |\calY| + |\calX|} \exp(-n \gamma), \label{eqn:use_lemma}
\end{align}
where  \eqref{eqn:union_bd} follows from the union bound and the fact that the codewords are generated in an identical  manner, and~\eqref{eqn:use_lemma} follows from Lemma~\ref{lem:conc_mi} noting that $\bX(2)$ is independent of $\bY$, the channel output when $\bX(1)$ is the input.

The erasure event can be expressed as
\begin{equation}
\calE_{\rme} := \calE_{\rme}^{(1)}\cup \calE_{\rme}^{(2)}
\end{equation}
where 
\begin{align}
\calE_{\rme}^{(1)} &:= \left\{  \hatI(\bX(m)\wedge\bY)  < \gamma ,\forall \, m\in [M_n ] \right\},\quad\mbox{and} \label{eqn:defE1} \\
\calE_{\rme}^{(2)} &:= \left\{  \hatI(\bX(m)\wedge\bY)  \ge \gamma , \mbox{for at least 2 messages } m \in [M_n]\right\} .
\end{align}
Clearly, we have that 
\begin{align}
\calE_{\rme}^{(1)} & \subset\calF_{\rme}^{(1)} :=  \left\{  \hatI(\bX(1)\wedge\bY)<\gamma\right\} , \quad\mbox{and} \label{eqn:defF1} \\
\calE_{\rme}^{(2)} & \subset\calF_{\rme}^{(2)} :=  \Big\{\max_{m \in [M_n]\setminus\{1\} } \hatI(\bX(m)\wedge\bY)\ge \gamma\Big\} .\label{eqn:defF2}
\end{align}
Note that the probability of $\calF_{\rme}^{(2)}$ was already bounded in \eqref{eqn:maxle}--\eqref{eqn:use_lemma}. Hence it remains to upper bound the probability of $\calF_{\rme}^{(1)}$ defined in~\eqref{eqn:defF1}. We let the random conditional type of $\bY$ given $\bX(1)$ be $U$.  Then, we have 
\begin{align}
\Pr\big[\calF_{\rme}^{(1)}\big]&  \le\Pr\big[ \hatI(\bX(1)\wedge\bY)\le\gamma\big] \\
&= \Pr\bigg[ I(P,W) + \sum_{x,y}( U(y|x)-W(y|x) )I_W'(y|x)   +O(\| U-W\|^2) \le\gamma\bigg] 
\end{align}
where the final step follows by Taylor expanding $U\mapsto I(P,U)$ around $U=W$ and  $I_W'(y|x) := \frac{\partial I(P,U)}{\partial U(y|x)}  \big|_{U=W}$. 
We also can bound the remainder term uniformly \cite{wang11} yielding
\begin{align}
\Pr\big[\calF_{\rme}^{(1)}\big] \le\Pr\bigg[ I(P,W) + \sum_{x,y}( U(y|x)-W(y|x) )I_W'(y|x)   \le\gamma + O\left(\frac{\log n}{n}\right)\bigg]  + O(n^{-2}) . \label{eqn:taylor_prob}
\end{align}
Wang-Ingber-Kochman~\cite{wang11} computed the relevant first-, second- and third-order statistics of the random variable $\sum_{x,y} (U(y|x) - W(y|x)) I_W'(y|x)$ allowing us to apply the Berry-Esseen theorem~\cite[Ch.~XVI.5]{feller} to the probability in  \eqref{eqn:taylor_prob}, leading to
\begin{align}
\Pr\big[\calF_{\rme}^{(1)}\big] \le \Phi\left( \frac{\gamma +  O(\frac{\log n}{n}) - I(P,W)}{\sqrt{V(P,W)/n}}\right)+ O(n^{-1/2})  . \label{eqn:berry}
\end{align}
Note that the implied constants in the $O(\fndot)$-notation in  \eqref{eqn:berry} are bounded because of the discreteness of the alphabets (e.g., \cite[Lem.~46]{PPV10}). Hence, we set 
\begin{equation}
\gamma=I(P,W)+ \sqrt{ \frac{V(P,W)}{n}}\Phi^{-1}(\ee')  
\end{equation}
to assert that 
\begin{equation}
 \Pr\big[\calF_{\rme}^{(1)}\big]  \le\ee' + O(n^{-1/2}).\label{eqn:erasure}
\end{equation}
We then set  $M_n$ to be the smallest integer satisfying
\begin{equation}
\log  M_n \ge n \gamma - \Big( |\calX||\calY|+|\calX | +\frac{1}{2}\Big)\log n.
\end{equation}
Then we may assert from \eqref{eqn:use_lemma} that 
\begin{equation}
\Pr [\calE_{\rmu}] \le   n^{-1/2}. \label{eqn:undetec}
\end{equation}
Furthermore, by the relations in  \eqref{eqn:defE1}--\eqref{eqn:defF2}, we have that 
\begin{equation}
\Pr [\calE_{\rme} ]  \le  \ee' + O(n^{-1/2}).\label{eqn:erasure2}
\end{equation}
Hence, we have 
\begin{equation}
\log  M_n  \ge  nI(P,W)  +  \sqrt{n  V(P,W) }\Phi^{-1}(\ee')    +   O\left(\log n\right) .
\end{equation}
Let $P_X^*$ achieve  $V_{\epsilon_{\rme}}$. This means that $P_X^* \in \argmin_{P\in\Pi}V(P,W)$ if $\epsilon_{\rme}<1/2$ and  $P_X^* \in \argmax_{P\in\Pi}V(P,W)$  if $\epsilon_{\rme}\ge 1/2$.
By choosing $P$ to be an $n$-type that is the  closest to $P_X^*$ (i.e., $P\in\argmin_{P\in\calP_n(\calX)}\| P-P_X^*\|_1$), we obtain, by the usual approximation arguments \cite[Lem.~7]{TomTan12},
\begin{equation}
\log  M_n\ge nC + \sqrt{n  V_{\ee'}  }\Phi^{-1}(\ee')   + O\left(\log n\right)  .\label{eqn:direct}
\end{equation}

We have proved the the random ensemble satisfies~\eqref{eqn:erasure} and~\eqref{eqn:undetec} but it is not clear yet there exists a  {\em deterministic code} that satisfies the same two bounds. To show this, let $\theta\in (0,1)$. Set $\eu:= \frac{1}{\theta } n^{-1/2}$ and $\ee := \frac{1}{ 1-\theta } (\ee' + O(n^{-1/2}))$ where $\ee' + O(n^{-1/2})$ denotes the right-hand-side of \eqref{eqn:erasure}. Then, making the  expectation over the random code $\calC$ explicit, \eqref{eqn:erasure} and \eqref{eqn:undetec} can be written as 
\begin{align}
\rmE \big[ \Pr[\calE_{\rme}|\calC]\big]    \le (1-\theta)\ee, \quad\mbox{and}\quad   \rmE\big[ \Pr[\calE_{\rmu}|\calC]\big]  \le \theta\eu.
\end{align}
Put $\eta :=\theta/2$. By Markov's inequality,
\begin{align}
\Pr [\calA ]  \le 1-\theta  ,\quad\mbox{and}\quad \Pr  [\calB]\le \theta - \eta, 
\end{align}
where  event  $\calA:= \{\Pr[\calE_{\rme}|\calC] >  \frac{1}{1-\theta} \rmE[ \Pr[\calE_{\rme}|\calC]] \}$ and event $\calB:=\{\Pr[\calE_{\rmu}|\calC] >  \frac{1}{\theta -\eta} \rmE[ \Pr[\calE_{\rmu}|\calC]]\}$. This implies that 
 there exists a {\em deterministic code} $\scC^*_n$ for which 
\begin{align}
 \Pr[\calE_{\rme}|\scC^*_n] & \le \frac{1}{1-\theta} \rmE[ \Pr[\calE_{\rme}|\calC]]\le\ee ,\quad\mbox{and} \\
 \Pr[\calE_{\rmu}|\scC^*_n]& \le \frac{1}{\theta-\eta} \rmE[ \Pr[\calE_{\rmu}|\calC]]\le  \frac{\theta}{\theta-\eta}  \eu= \frac{2}{\theta \sqrt{n}}.  \label{eqn:deterministic}
\end{align}
Letting $n\to\infty$, we conclude there exists a sequence of deterministic codes $\{ \scC^*_n\}_{n\ge 1}$ such that 
\begin{equation}
\limsup_{n\to\infty}\Pr[\calE_{\rme}|\scC^*_n] \le \ee,\quad\mbox{and}\quad  \lim_{n\to\infty}\Pr[\calE_{\rmu}|\scC^*_n] = 0.
\end{equation}
Now, according to~\eqref{eqn:direct} and the relation between $\ee$ and $\ee'$, 
\begin{align}
 \liminf_{n\to\infty}\frac{1}{\sqrt{n}} (\log  M_n  -  nC)  \ge  \sqrt{V_{(1 - \theta)\ee}} \Phi^{-1}((1 -    \theta)\ee) .
\end{align}
Finally  take $\theta\downarrow 0$ to complete the proof.
\subsection{List Decoding: Proof of Theorem~\ref{thm:list}} \label{sec:prf_list} 
We first prove the non-asymptotic converse bound for list decoding (Proposition \ref{prop:list}) in Sec.~\ref{sec:prf_prop}. Subsequently, we prove   \eqref{eqn:r_second}--\eqref{eqn:s_third2}  in unison in both the converse (Sec.~\ref{sec:converse_list_prf}) and direct  (Sec.~\ref{sec:prf_list_direct}) parts.
\subsubsection{Proof of Proposition~\ref{prop:list}} \label{sec:prf_prop}
We modify the argument leading to  \cite[Prop.~6]{TomTan12}, which was inspired by \cite{Wang09,WangRenner}, so that it is applicable to the list decoding setting.  Note that this result was also proven by Kostina-Verd\'u~\cite[Thm.~4]{kost13} for the joint source-channel  setting. Let $P$ and $Q$ be two  probability measures on the same space $\calZ$. Let $\delta:\calZ\to \{0,1\}$ represent a (deterministic) hypothesis test between $P$ and $Q$ where $\delta=1$ implies deciding in favor  of $P$.  We define the {\em $\epsilon$-hypothesis testing divergence}
\begin{align}
D_{\rmh}^\epsilon(P\|Q) & := \sup\Big\{ R\in\bbR : \exists\, \delta:\calZ \to \{0,1\},  \rmE_Q [\delta(Z)]\le(1-\epsilon)\exp(-R) ,  \nn\\
 &\qquad\qquad\qquad  \,\,\,\,\qquad\qquad\qquad\qquad\rmE_P[\delta(Z)]\ge 1-\epsilon\Big\}  \label{eqn:Dh_def} 
\end{align}
Note that the $\epsilon$-hypothesis testing divergence is related to $\beta_{1-\epsilon}$ as follows:
\begin{equation}
D_{\rmh}^\epsilon(P\|Q)= -\log\frac{ \beta_{1-\epsilon}(P,Q)}{ 1-\epsilon}.
\end{equation}
Some properties of $D_{\rmh}^\epsilon$ include (i)  $D_{\rmh}^\epsilon(P\|Q)\ge 0$ with equality iff $P=Q$ almost everywhere \cite[Prop.~3.2]{Dup12}; (ii) For any random transformation $U:\calZ\to\calZ'$, the data processing inequality~\cite[Prop.~6]{TomTan12}  implies that 
\begin{equation}
D_{\rmh}^\epsilon  (PU\| QU)\le D_{\rmh}^\epsilon  (P\| Q) . \label{eqn:dpi}
\end{equation}

Any $(|\calM|, L, \epsilon)_{\mathrm{ave}}$-list code $(f,\varphi)$ with message set $\calM$ (Definition~\ref{def:list_codes})  induces the Markov chain $M - X-Y-\calS_L$ where $M$ is uniform on $\calM$ and $\calS_L \in \binom{\calM}{\le L}$ is a random subset of $\calM$ with size not exceeding $L$. This defines the joint distribution 
\begin{equation}
P_{MXY\calS_L} (m, x, y, s_L) = \frac{1}{|\calM|}\bone\{x=f(m)\} W(y|x) \bone\{\varphi(m)\in s_L\}.
\end{equation}
Fix $Q\in\calP(\calY)$. Due to the data processing inequality for $D_{\rmh}^\epsilon$ in \eqref{eqn:dpi}, we have 
\begin{equation}
D_{\rmh}^\epsilon (P\times W \| P\times Q) = D_{\rmh}^\epsilon (P_{XY} \| P_X \times Q_Y)\ge  D_{\rmh}^\epsilon (P_{M \calS_L} \| P_M \times Q_{\calS_L}) \label{eqn:dp}
\end{equation}
where $P_X=P$ and $Q_{\calS_L}$ is the distribution on the subsets of $\calM$ of size not exceeding $L$ induced by the decoder applied to $Q_Y=Q$. Moreover, consider the test (with the identification of the space $\calZ\leftarrow\calM\times  \binom{\calM}{\le L}$  in \eqref{eqn:Dh_def})
\begin{equation}
\delta(M, \calS_L) := \bone\{ M\in\calS_L\}. 
\end{equation}
Then under the null hypothesis  $P_{M\calS_L}$, we have 
\begin{align}
\rmE_{P_{M \calS_L}}[\delta(M, \calS_L) ]  = \Pr[ M \in \calS_L ] \ge 1-\epsilon
\end{align}
because the average error probability   of the list code is no larger than $\epsilon$. Furthermore, under the alternate hypothesis $P_M \times Q_{\calS_L}$,
\begin{align}
\rmE_{P_M\times Q_{\calS_L}}[ \delta(M, \calS_L)]  &= \sum_{m\in\calM} \frac{1}{|\calM|} \sum_{\scS \in \binom{\calM}{\le L}} Q_{\calS_L}(\scS)\delta(m,\scS) \\
&  = \frac{1}{|\calM|}  \sum_{\scS\in \binom{\calM}{\le L}} Q_{\calS_L}(\scS) \sum_{m\in\calM} \delta(m,\scS) \\
&  =  \frac{1}{|\calM|}  \sum_{\scS\in \binom{\calM}{\le L}} Q_{\calS_L}(\scS) |\scS| \\
&  \le  \frac{L}{|\calM|}   \sum_{\scS\in \binom{\calM}{\le L}} Q_{\calS_L}(\scS)   \label{eqn:lessL} \\
&  = \frac{L}{|\calM|},
\end{align}
where \eqref{eqn:lessL} follows because $|\scS|\le L$ for all subsets $\scS \in \binom{\calM}{\le L}$.  Hence, by the definition of the $\epsilon$-hypothesis testing divergence $D_{\rmh}^\epsilon$ in \eqref{eqn:Dh_def}, we have 
\begin{equation}
D_{\rmh}^\epsilon (P_{M \calS_L} \| P_M \times Q_{\calS_L})\ge  \log \frac{ |\calM |}{L}+\log (1-\epsilon). \label{eqn:def_dh}
\end{equation}
By uniting \eqref{eqn:dp} and \eqref{eqn:def_dh}, maximizing over $P\in\calP(\calX)$ to make the bound code independent, and minimizing  over $Q \in\calP(\calY)$ which was arbitrary, we obtain
\begin{equation}
\log \frac{|\calM|}{L}\le \inf_{Q \in\calP(\calY)}\sup_{P\in\calP(\calX)} D_{\rmh}^{\epsilon} (P\times W   \|  P\times Q )  +\log\frac{1}{1-\epsilon}\label{eqn:gleaned}
\end{equation}
which is exactly the same as \eqref{eqn:meta_list} in Proposition~\ref{prop:list}.

\subsubsection{Converse Part} \label{sec:converse_list_prf} 
We assume that  $V_\epsilon>0$. Now, we may choose  $Q_{\bY} \in \calP(\calY^n)$ as in \cite[Sec III.C]{TomTan12} and follow the analysis in  \cite[Props.~6,  8 and~10(i)]{TomTan12} to upper bound $D_{\rmh}^{\epsilon} (P_{\bX}\times W^n \|P_{\bX}\times  Q_{\bY})$ by relaxing it to a quantity known as the $\epsilon$-information spectrum divergence. This yields 
\begin{equation}
\log \frac{M_n}{L_n}\le nC+\sqrt{nV_\epsilon}\Phi^{-1}(\epsilon)+\frac{1}{2}\log n + O(1).
\end{equation}
To prove  \eqref{eqn:r_second}, notice that $\log L_n =  O(\sqrt{n}\, l)$ so the second-order term $r^*_{\mathrm{list}, \rma}(l,\epsilon)\le l+ \sqrt{n V_\epsilon}\Phi^{-1}(\epsilon)$. For \eqref{eqn:s_third}, $L_n =  O(n^{\alpha})$, so the third-order term $s^*_{\mathrm{list}}(\alpha,\epsilon)\le \frac{1}{2}+\alpha$ as desired.

 To prove the stronger statement concerning symmetric and singular DMCs in \eqref{eqn:s_third2}, we note that per~\cite[Thm.~22]{Pol13},  the saddle-point in \eqref{eqn:gleaned} is attained for $P_{\bX}$ being the uniform distribution on $\calX^n$. We also use the output distribution $Q_{\bY}^{(\mathrm{AW}) }$ suggested    by Altu\u{g} and Wagner in \cite[Sec.~IV.A]{altug13}. This shows that $ D_{\rmh}^{\epsilon } (P_{\bX}\times W^n  \| P_{\bX}\times  Q_{\bY}^{(\mathrm{AW} )})$ is bounded above by the Gaussian approximation plus  a constant term, completing the proof of $s^*_{\mathrm{list}}(\alpha,\epsilon)\le \alpha$ in view of the fact that $L_n= O(n^\alpha)$. 

\subsubsection{Direct Part}  \label{sec:prf_list_direct}
To get the   third-order result for the lower bound in \eqref{eqn:s_third}, we use i.i.d.\ random codes with distribution $P^*_X$ achieving $V_\epsilon$. We also use threshold decoding of the information density. Contrast this to using constant composition codes and maximum empirical mutual information decoding in Sec.~\ref{sec:era_prf_dir} which generally results in worse third-order terms. Now the decoder outputs {\em all}  messages $m$ whose log-likelihood ratio exceeds  $\gamma\ge 0$, i.e.\  the list is 
\begin{equation}
\calL:=\left\{m \in [M_n]: \frac{1}{n}\log\frac{W^n(\bY|\bX(m))}{(P^*_X W)^n(\bY)}\ge\gamma \right\}.
\end{equation}
We now analyze the error probability assuming that message $m=1$ was sent. The two error events are 
\begin{align}
\calE_1  &:= \left\{ \frac{1}{n}\log\frac{W^n(\bY|\bX(1))}{(P^*_X W)^n(\bY)} < \gamma\right\} ,\quad\mbox{and}\\
\calE_2 & := \left\{ |\calL|>  L_n \right\} . 
\end{align}
The probability of $\calE_1$ can be analyzed using the Berry-Esseen theorem~\cite[Ch.~XVI.5]{feller}. This yields
\begin{equation}
\Pr[\calE_1] \le \Phi\left(\frac{\gamma-C}{\sqrt{V_\epsilon/n}}\right) + O(n^{-1/2}). \label{eqn:err1}
\end{equation}
The $\epsilon$-dispersion $V_\epsilon$ appears in the denominator in \eqref{eqn:err1} because the unconditional information variance equals the conditional information variance for capacity-achieving distributions~\cite[Lem.~62]{PPV10} and $P^*_X$ is chosen to achieve $V_\epsilon$.  Now, consider the expectation of the size of the list of incorrect messages $\calL_{\setminus 1} := \calL \setminus \{1\}$. Indeed
\begin{align}
\rmE \big[|\calL_{\setminus 1} |\big]  &= \rmE \left[\sum_{m =2}^{M_n} \bone \bigg\{  \frac{1}{n}\log\frac{W^n(\bY|\bX(m))}{(P^*_XW)^n(\bY)}\ge\gamma\bigg\} \right]\\
&=  \sum_{m =2}^{M_n} \Pr\bigg[ \frac{1}{n}\log\frac{W^n(\bY|\bX(m))}{(P^*_X W)^n(\bY)}\ge\gamma  \bigg] .\label{eqn:probs}
\end{align}
Now, we analyze the probability in \eqref{eqn:probs}.
By symmetry,   all summands in  \eqref{eqn:probs} are identical so we just focus on the $m=2$ term. By using a strong large-deviations result due to Polyanskiy-Poor-Verd\'u~\cite[Lem.~47]{PPV10},  and a standard change-of-measure argument, for every $n\in\bbN$, we have
\begin{equation}
\Pr\bigg[ \frac{1}{n}\log\frac{W^n(\bY|\bX(2))}{(P^*_X W)^n(\bY)}\ge\gamma  \bigg] \le  \frac{a\, \exp(-n\gamma) }{\sqrt{n}}  \label{eqn:prob_rootn} ,
\end{equation}
where $a=a_W>0$ is a  finite channel-dependent parameter.
 Here we made use of the fact that $V_\epsilon>0$ and the third absolute moment of the information density for discrete memoryless systems is finite. 
The result in \eqref{eqn:prob_rootn} can also be obtained from~\cite[Prop.~6.2]{mou13b}, which follows from a strong large-deviations result of Chaganty and  Sethuraman \cite[Thms.~3.3~and~3.5]{CS93} but \eqref{eqn:prob_rootn} as stated is non-asymptotic. 
Thus, plugging \eqref{eqn:prob_rootn} into  \eqref{eqn:probs} and bounding $M_n-1$ by $M_n$ we have that 
\begin{equation}
\rmE \big[|\calL_{\setminus 1}|\big] \le  \frac{M_n b\exp(-n\gamma)}{\sqrt{n}} \label{eqn:exp_list}
\end{equation}
for some channel dependent constant $b>0$ and $n$ sufficiently large. 
Since $|\calL| \le |\calL_{\setminus 1} | + 1$ (consider the two different cases in which $1\in \calL$ and otherwise),  the probability of $\calE_2$  can be bounded as 
\begin{equation}
\Pr[\calE_2] = \Pr\big[ |\calL | > L_n \big] \le  \Pr\big[ |\calL_{\setminus 1} |  > L_n-1 \big].
\end{equation}
By Markov's inequality,
\begin{equation}
\Pr[\calE_2] \le   \frac{\rmE \big[|\calL_{\setminus 1}|\big]  }{L_n-1}\le  \frac{M_n b\exp(-n\gamma)}{\sqrt{n} \, (L_n-1)}. \label{eqn:probs2}
\end{equation}
Now choose 
\begin{equation}
\gamma = C +\sqrt{\frac{V_\epsilon}{n}} \Phi^{-1}(\epsilon) \label{eqn:delta_n}
\end{equation}
so $\Pr[\calE_1]\le\epsilon+ O(n^{-1/2})$ according to~\eqref{eqn:err1}.  Also choose 
\begin{equation}
M_n = \lfloor(L_n -1)\exp(n\gamma)\rfloor \label{eqn:choice_M}
\end{equation}
so $\Pr[\calE_2|\calE_1^c] \le  O(n^{-1/2})$ according to~\eqref{eqn:probs2}.  As such the total error probability $\Pr[\calE_1\cup\calE_2]$ (probability that the true message does not belong to the list or the list size  exceeds $L_n$) is bounded from above by $\epsilon+  O(n^{-1/2})$, which  satisfies (the more stringent condition in) \eqref{eqn:error_rootn}. 

In the second-order case in which $\log L_n = O(\sqrt{n}\, l)$,  the choice of $M_n$ in \eqref{eqn:choice_M} yields $r^*_{\mathrm{list}, \rma}(l,\epsilon)\ge l+ \sqrt{n V_\epsilon}\Phi^{-1}(\epsilon)$, proving the direct part of~\eqref{eqn:r_second}. In the third-order case in which $L_n = O(n^\alpha)$,  the choice of $M_n$ in~\eqref{eqn:choice_M}  yields $s^*_{\mathrm{list}}(\alpha,\epsilon)\ge \alpha$, proving the direct part of~\eqref{eqn:s_third}. 

Note that unlike the erasures setting, in this case we do not need to augment the proof with the argument involving Markov's inequality (cf.~argument leading to \eqref{eqn:deterministic}) because here, there is only a single error criterion.

\subsection{Slepian-Wolf with Erasure Option: Proof of Theorem~\ref{thm:sw}} \label{sec:prf_sw}
\subsubsection{Converse part} The technique developed by Miyake and Kanaya~\cite[Sec.~4.2]{miyake} applies directly. This allows us to conclude that every $(M_n, \eun, \etn)$-code for the correlated source must satisfy
\begin{equation}
\etn \ge \Pr\left(\frac{1}{n}\log\frac{1}{P_{\bX|\bY}(\bX|\bY)}\ge\frac{1}{n}\log M_n+\gamma \right)+\exp(-n\gamma), \label{eqn:miyake}
\end{equation}
for any $\gamma>0$ regardless of $\eun\in [0,\etn)$. We  immediately  obtain the converse by setting $\gamma=\frac{\log n}{2n}$ and  applying the Berry-Esseen theorem~\cite[Ch.~XVI.5]{feller} to the probability in \eqref{eqn:miyake}. See the proof of the converse of~\cite[Thm.~1]{TK14} for details.

\subsubsection{Direct part} By the same argument as the channel coding case in Sec.~\ref{sec:era_prf_dir}, it suffices to show that $\sqrt{V} \Phi^{-1}(1-\et)$ is $(0,\et)$-achievable. 

As in Sec.~\ref{sec:era_prf_dir}, we  provide a universal coding scheme. Randomly and independently partition the set of source sequences $\calX^n$ into $M_n=\lceil\exp(nR_n)\rceil$ bins where  the ``rate'' $R_n$ is to be chosen later. Let $\calB(m)$ be the set of source sequences $\bx$ which are (randomly) mapped to the bin  indexed by $m \in [M_n]$. If $\bx$ is received by the encoder, send message $m$. 

Let $\hatH(\bx|\by)$ be the empirical conditional entropy of the sequences $(\bx,\by)$, i.e., $\hatH(\bx|\by)=H(U|P_{\by})$ where $U$ is the conditional type of $\bx$ given $\by$. The decoder, given $\by \in \calY^n$ and $m \in [M_n]$, finds a unique source sequence $\hat{\bx}$ in $\calB(m)$ satisfying
\begin{equation}
\hatH(\hat{\bx} | \by) \le\gamma
\end{equation}
for some threshold $\gamma$, which will be chosen later. If there is no such source sequence in bin $\calB(m)$ or there is more than one, declare an erasure event. 

Let $\bX$ and $\bY$ be the randomly generated sequences from $P_{XY}^n(\bx,\by):=\prod_{j=1}^n P_{XY}(x_j, y_j)$.  The probability of undetected error can be bounded as 
\begin{align}
\Pr[\calE_{\rmu}] \le \Pr\left[ \exists \, \tilde{\bx} \in\calB(M) \setminus\{\bX\} :  \hatH(\tilde{\bx}|\bY)  \le \gamma \right], \label{eqn:undet}
\end{align}
where $M$ is the bin index corresponding to the random source $\bX$. By symmetry, it is enough to fix the bin index to be $M=1$. Now, we condition on various values of $(\bx,\by ) \in\calX^n\times\calY^n$ as follows:
\begin{align}
\Pr[\calE_{\rmu}] & \le\sum_{\bx,\by} P_{XY}^n(\bx,\by) \Pr\left[ \exists \, \tilde{\bx} \in\calB(1) \setminus\{\bx\} :  \hatH(\tilde{\bx}|\bY)  \le \gamma \,\Big|\,\bX=\bx,\bY=\by \right] \\
&\le \sum_{\bx,\by} P_{XY}^n(\bx,\by) \sum_{\tilde{\bx} \ne \bx: \hatH(\tilde{\bx}|\by) \le \gamma} \Pr\left[   \tilde{\bx} \in\calB(1)  \right] \label{eqn:binning_indep}\\
&= \sum_{\bx,\by} P_{XY}^n(\bx,\by) \sum_{U\in\calU_n(\calX; P_{\by}) : H(U|P_{ \by}) \le \gamma} \sum_{\tilde{\bx} \in\calT_{U}(\by)} \Pr\left[   \tilde{\bx} \in\calB(1)  \right]  \label{eqn:split_cond_types}\\
&\le \sum_{\bx,\by} P_{XY}^n(\bx,\by) \sum_{U\in\calU_n(\calX; P_{\by}) : H(U|P_{ \by}) \le \gamma} |\calT_{U}(\by)| \exp(-nR_n) \label{eqn:uniform}\\
&\le \sum_{\bx,\by} P_{XY}^n(\bx,\by) \sum_{U\in\calU_n(\calX; P_{\by}) : H(U|P_{ \by}) \le \gamma} \exp(nH(U|P_{\by})) \exp(-nR_n) \label{eqn:size_of}\\
&\le \sum_{\bx,\by} P_{XY}^n(\bx,\by) \sum_{U\in\calU_n(\calX; P_{\by}) : H(U|P_{ \by}) \le \gamma} \exp(n\gamma) \exp(-nR_n) \label{eqn:use_clause}\\
&\le (n+1)^{|\calX| |\calY|} \exp(-n (R_n-\gamma)),  \label{eqn:type_counting}
\end{align}
where \eqref{eqn:binning_indep} follows from the fact that the binning is independent of the generation of $\bx,\by$, in \eqref{eqn:split_cond_types} we partitioned $\tilde{\bx}$ into $U$-shells compatible with the type of $\by$, \eqref{eqn:uniform} follows from the uniformity in binning and the fact that the number of bins is $M_n=\lceil\exp( nR_n)\rceil$, \eqref{eqn:size_of} follows from the fact that $|\calT_U(\by) | \le \exp(n H(U | P_{\by}))$, \eqref{eqn:use_clause} uses the fact that $H(U | P_{\by})\le  \gamma$, and finally, \eqref{eqn:type_counting} follows from the type counting lemma \cite[Lem.~2.1]{Csi97}.

Let the erasure event be $\calE_{\rme}$. Then, $\calE_{\rme}$ is the union of $\calE_{\rme}^{(1)}$, the event that all $\bx\in\calB(M)$ satisfy $\hatH(\bx|\bY)>\gamma$ and $\calE_{\rme}^{(2)}$, the event that there are $2$ or more sequences $\bx_j \in\calB(M),j\in\calJ$ (with $|\calJ|\ge 2$) such that   $\hatH(\bx_j|\bY)\le\gamma$ for all $j\in\calJ$. As mentioned above, by symmetry, we may assume $M=1$. Then, clearly, 
\begin{align}
 \calE_{\rme}^{(1)} \subset\calF_{\rme}^{(1)}  &:= \big\{ \hatH(\bX|\bY)  > \gamma   \big\}  \label{eqn:defF1_sw}\\
  \calE_{\rme}^{(2)} \subset\calF_{\rme}^{(2)} &:= \big\{\exists \, \tilde{\bx} \in\calB(1) \setminus\{\bX\} :  \hatH(\tilde{\bx}|\bY)  \le \gamma\big\}  \label{eqn:defF2_sw}
\end{align}
The probability of $\calF_{\rme}^{(2)}$ was already bounded in the steps leading to \eqref{eqn:type_counting} so it remains to bound the probability of $\calF_{\rme}^{(1)}$. We have 
\begin{align}
\Pr\big[\calF_{\rme}^{(1)} \big]&  = \Pr\big[ \hatH(\bX|\bY)  > \gamma \big]  \\
 &\le \Phi\left( \frac{H(X|Y)-\gamma}{\sqrt{V(X|Y)/n}}\right) + O(n^{-1/2}), \label{eqn:eras_sw}
\end{align}
where the last step follows by Taylor expanding $\hatH(\bX|\bY)$ around $H(X|Y)$ and then applying the Berry-Esseen theorem~\cite[Ch.~XVI.5]{feller}. See the so-called {\em vector rate redundancy theorem} in~\cite{TK14} for details.  

We now choose 
\begin{equation}
\gamma = H(X|Y) + \sqrt{ \frac{V(X|Y)}{n}} \Phi^{-1}(1-\et) \label{eqn:def_gam}
\end{equation}
 so the probability of $\calF_{\rme}^{(1)}$ in \eqref{eqn:eras_sw} is upper bounded by $\et +O(n^{-1/2})$. We also pick 
\begin{equation}
 R_n = \gamma + (|\calX||\calY| + 1/2)\frac{\log (n+1)}{n} \label{eqn:def_Rn}
\end{equation}
 so the probability of undetected error in~\eqref{eqn:type_counting} is upper bounded by $O(n^{-1/2})$.  By \eqref{eqn:defF1_sw}--\eqref{eqn:defF2_sw} and the preceding conclusions, these choices also yield
\begin{equation}
\Pr[\calE_{\rme}]\le \et +O(n^{-1/2}).
 \end{equation} 
Now we employ    the Markov inequality argument at the end of the proof in Section~\ref{sec:era_prf_dir}. With this and the realization that 
\begin{equation}
\limsup_{n\to\infty}\frac{1}{\sqrt{n}}(\log M_n-nH(X|Y)) \le \sqrt{V(X|Y)} \Phi^{-1}(1-\et)
\end{equation}
from \eqref{eqn:def_gam}--\eqref{eqn:def_Rn}, we have   proved that  $\sqrt{V(X|Y)}\Phi^{-1}(1-\et)$ is a $(0,\et)$-achievable second-order coding rate for the Slepian-Wolf problem with correlated source $P_{XY}$.

\subsubsection*{Acknowledgements}
The authors would like to thank  Yanina Shkel for sharing her  MATLAB code to generate the finite blocklength performance bounds in Fig.~\ref{fig:eras}. He would also like to  acknowledge helpful discussions with Wei Yang, Masahito Hayashi, Marco Tomamichel and Jonathan Scarlett. 
\bibliographystyle{unsrt}
\bibliography{isitbib}

\end{document}

%% file: preamble.tex
\usepackage[mathscr]{eucal}
\usepackage[cmex10]{amsmath}
\usepackage{epsfig,epsf,psfrag,overpic}
\usepackage{amssymb,amsmath,amsthm,amsfonts,latexsym}
\usepackage{amsmath,graphicx,bm,xcolor,url}
\usepackage[caption=false]{subfig} 
\usepackage{fixltx2e}
\usepackage{array}
\usepackage{verbatim}
\usepackage{bm}
\usepackage{algorithmic, cite}
\usepackage{algorithm}
\usepackage{verbatim}
\usepackage{textcomp}
\usepackage{mathrsfs}
\usepackage{epstopdf}

\catcode`~=11 \def\UrlSpecials{\do\~{\kern -.15em\lower .7ex\hbox{~}\kern .04em}} \catcode`~=13 

\allowdisplaybreaks[1]
 
\newcommand{\nn}{\nonumber}

\newcommand{\calA}{\mathcal{A}}
\newcommand{\calB}{\mathcal{B}}
\newcommand{\calC}{\mathcal{C}}
\newcommand{\calD}{\mathcal{D}}
\newcommand{\calE}{\mathcal{E}}
\newcommand{\calF}{\mathcal{F}}

\newcommand{\calJ}{\mathcal{J}}

\newcommand{\calL}{\mathcal{L}}
\newcommand{\calM}{\mathcal{M}}

\newcommand{\calP}{\mathcal{P}}

\newcommand{\calS}{\mathcal{S}}
\newcommand{\calT}{\mathcal{T}}
\newcommand{\calU}{\mathcal{U}}

\newcommand{\calX}{\mathcal{X}}
\newcommand{\calY}{\mathcal{Y}}
\newcommand{\calZ}{\mathcal{Z}}

\newcommand{\bx}{\mathbf{x}}
\newcommand{\bX}{\mathbf{X}}
\newcommand{\by}{\mathbf{y}}
\newcommand{\bY}{\mathbf{Y}}

\newcommand{\rma}{\mathrm{a}}

\newcommand{\rmb}{\mathrm{b}}

\newcommand{\rmc}{\mathrm{c}}

\newcommand{\rmd}{\mathrm{d}}

\newcommand{\rme}{\mathrm{e}}
\newcommand{\rmE}{\mathrm{E}}

\newcommand{\rmh}{\mathrm{h}}

\newcommand{\rmt}{\mathrm{t}}

\newcommand{\rmu}{\mathrm{u}}

\newcommand{\bbN}{\mathbb{N}}

\newcommand{\bbR}{\mathbb{R}}

\newcommand{\scC}{\mathscr{C}}

\newcommand{\scS}{\mathscr{S}}

\DeclareMathAlphabet{\mathbsf}{OT1}{cmss}{bx}{n}
\DeclareMathAlphabet{\mathssf}{OT1}{cmss}{m}{sl}

\newcommand{\rvp}{\mathsf{p}}

\DeclareSymbolFont{bsfletters}{OT1}{cmss}{bx}{n}  
\DeclareSymbolFont{ssfletters}{OT1}{cmss}{m}{n}
\DeclareMathSymbol{\bsfGamma}{0}{bsfletters}{'000}
\DeclareMathSymbol{\ssfGamma}{0}{ssfletters}{'000}
\DeclareMathSymbol{\bsfDelta}{0}{bsfletters}{'001}
\DeclareMathSymbol{\ssfDelta}{0}{ssfletters}{'001}
\DeclareMathSymbol{\bsfTheta}{0}{bsfletters}{'002}
\DeclareMathSymbol{\ssfTheta}{0}{ssfletters}{'002}
\DeclareMathSymbol{\bsfLambda}{0}{bsfletters}{'003}
\DeclareMathSymbol{\ssfLambda}{0}{ssfletters}{'003}
\DeclareMathSymbol{\bsfXi}{0}{bsfletters}{'004}
\DeclareMathSymbol{\ssfXi}{0}{ssfletters}{'004}
\DeclareMathSymbol{\bsfPi}{0}{bsfletters}{'005}
\DeclareMathSymbol{\ssfPi}{0}{ssfletters}{'005}
\DeclareMathSymbol{\bsfSigma}{0}{bsfletters}{'006}
\DeclareMathSymbol{\ssfSigma}{0}{ssfletters}{'006}
\DeclareMathSymbol{\bsfUpsilon}{0}{bsfletters}{'007}
\DeclareMathSymbol{\ssfUpsilon}{0}{ssfletters}{'007}
\DeclareMathSymbol{\bsfPhi}{0}{bsfletters}{'010}
\DeclareMathSymbol{\ssfPhi}{0}{ssfletters}{'010}
\DeclareMathSymbol{\bsfPsi}{0}{bsfletters}{'011}
\DeclareMathSymbol{\ssfPsi}{0}{ssfletters}{'011}
\DeclareMathSymbol{\bsfOmega}{0}{bsfletters}{'012}
\DeclareMathSymbol{\ssfOmega}{0}{ssfletters}{'012}

\newcommand{\hatH}{\hat{H}}

\newcommand{\hatI}{\hat{I}}

\newcommand{\hatm}{\hat{m}}

\newcommand{\tilm}{\tilde{m}}

\newcommand{\tilX}{\tilde{X}}

\newcommand{\tilY}{\tilde{Y}}

\newcommand{\barx}{\bar{x}}

\def\fndot{\, \cdot \,}

\DeclareMathOperator*{\argmax}{arg\,max}
\DeclareMathOperator*{\argmin}{arg\,min}

\DeclareMathOperator{\var}{\mathsf{Var}}

\newcommand{\bone}{\mathbf{1}}

\newtheorem{theorem}{Theorem} 
\newtheorem{lemma}[theorem]{Lemma}

\newtheorem{proposition}[theorem]{Proposition}

\newtheorem{definition}{Definition}

\newcommand{\qednew}{\nobreak \ifvmode \relax \else
      \ifdim\lastskip<1.5em \hskip-\lastskip
      \hskip1.5em plus0em minus0.5em \fi \nobreak
      \vrule height0.75em width0.5em depth0.25em\fi}

%% file: erasures_extended.bbl
\begin{thebibliography}{10}

\bibitem{Forney68}
G.~D. Forney.
\newblock Exponential error bounds for erasure, list, and decision feedback
  schemes.
\newblock {\em IEEE Trans. on Inf. Th.}, 14:206--220, 1968.

\bibitem{sgb}
C.~E. Shannon, R.~G. Gallager, and E.~R. Berlekamp.
\newblock Lower bounds to error probability for coding in discrete memoryless
  channels {I-II}.
\newblock {\em Information and Control}, 10:65--103,522--552, 1967.

\bibitem{tel94}
I.~E. Telatar and R.~G. Gallager.
\newblock {New exponential upper bounds to error and erasure probabilities}.
\newblock In {\em Intl.\ Symp.\ on Inf.\ Th.}, page 379, Trondheim, Norway,
  1994.

\bibitem{Blinovsky}
V.~M. Blinovsky.
\newblock Error probability exponent of list decoding at low rates.
\newblock {\em Problems of Information Transmission}, 27(4):277--287, 2001.

\bibitem{Moulin09}
P.~Moulin.
\newblock A {Neyman-Pearson} approach to universal erasure and list decoding.
\newblock {\em IEEE Trans. on Inf. Th.}, 55:4462--4478, Oct 2009.

\bibitem{merhav08}
N.~Merhav.
\newblock Error exponents of erasure/list decoding revisited via moments of
  distance enumerators.
\newblock {\em IEEE Trans. on Inf. Th.}, 54:4439--4447, Oct 2008.

\bibitem{merhav13}
N.~Merhav.
\newblock List decoding--random coding exponents and expurgated exponents.
\newblock {\em submitted to the IEEE Trans. on Inf. Th.}, 2013.
\newblock arXiv:1311.7298.

\bibitem{Strassen}
V.~Strassen.
\newblock {Asymptotische Absch\"{a}tzungen in Shannons Informationstheorie}.
\newblock In {\em Trans. Third Prague Conf. Inf. Theory}, pages 689--723,
  Prague, 1962.
\newblock Available at http://www.math.cornell.edu/$\sim$pmlut/strassen.pdf.

\bibitem{Kemperman}
J.~H.~B. Kemperman.
\newblock {Studies in Coding Theory I}.
\newblock Technical report, University of Rochester, NY, 1962.
\newblock Available at https://www.dropbox.com/s/amrio27pea0vz1f/Kemperman.pdf.

\bibitem{PPV10}
Y.~Polyanskiy, H.~V. Poor, and S.~Verd\'{u}.
\newblock Channel coding rate in the finite blocklength regime.
\newblock {\em IEEE Trans. on Inf. Th.}, 56:2307--2359, May 2010.

\bibitem{Hayashi09}
M.~Hayashi.
\newblock Information spectrum approach to second-order coding rate in channel
  coding.
\newblock {\em IEEE Trans. on Inf. Th.}, 55:4947--4966, Nov 2009.

\bibitem{mou13b}
P.~Moulin.
\newblock The log-volume of optimal codes for memoryless channels, within a few
  nats.
\newblock Nov 2013. arXiv:1311.0181.

\bibitem{TomTan12}
M.~Tomamichel and V.~Y.~F. Tan.
\newblock A tight upper bound for the third-order asymptotics of most discrete
  memoryless channels.
\newblock {\em IEEE Trans. on Inf. Th.}, 59(11):7041--7051, Nov 2013.

\bibitem{altug13}
Y.~Altu\u{g} and A.~Wagner.
\newblock The third-order term in the normal approximation for singular
  channels.
\newblock 2013. arXiv:309.5126.

\bibitem{TanTom13}
V.~Y.~F. Tan and M.~Tomamichel.
\newblock The third-order term in the normal approximation for the {AWGN}
  channel.
\newblock 2013. arXiv:1311.2237v2.

\bibitem{TK14}
V.~Y.~F. Tan and O.~Kosut.
\newblock On the dispersions of three network information theory problems.
\newblock {\em IEEE Trans. on Inf. Th.}, 60(2):881--903, Feb 2014.

\bibitem{sw73}
D.~Slepian and J.~K. Wolf.
\newblock Noiseless coding of correlated information sources.
\newblock {\em IEEE Trans. on Inf. Th.}, 19:471--80, 1973.

\bibitem{gallagerIT}
R.~G. Gallager.
\newblock {\em {Information Theory and Reliable Communication}}.
\newblock Wiley, New York, 1968.

\bibitem{Csi97}
I.~Csisz\'{a}r and J.~{K\"{o}rner}.
\newblock {\em Information Theory: Coding Theorems for Discrete Memoryless
  Systems}.
\newblock Cambridge University Press, 2011.

\bibitem{merhav13a}
N.~Merhav.
\newblock Erasure/list exponents for {Slepian-Wolf} decoding.
\newblock {\em submitted to the IEEE Trans. on Inf. Th.}, 2013.
\newblock arXiv:1305.5626.

\bibitem{PPV11b}
Y.~Polyanskiy, H.~V. Poor, and S.~Verd\'{u}.
\newblock Feedback in the non-asymptotic regime.
\newblock {\em IEEE Trans. on Inf. Th.}, 57(8):4903--4925, 2011.

\bibitem{chen13}
T.-Y. Chen, A.~R. Williamson, N.~Seshadri, and R.~D. Wesel.
\newblock Feedback communication systems with limitations on incremental
  redundancy.
\newblock {\em submitted to the IEEE Trans. on Inf. Th.}, 2013.
\newblock arXiv:1309.0707.

\bibitem{Will13}
A.~R. Williamson, T.-Y. Chen, and R.~D. Wesel.
\newblock Reliability-based error detection for feedback communication with low
  latency.
\newblock In {\em Intl.\ Symp.\ Inf.\ Th.}, Istanbul, Turkey, 2013.

\bibitem{Pol10}
Y.~Polyanskiy.
\newblock {\em Channel coding: {Non}-asymptotic fundamental limits}.
\newblock PhD thesis, Princeton University, 2010.

\bibitem{kost13}
V.~Kostina and S.~Verd\'{u}.
\newblock Lossy joint source-channel coding in the finite blocklength regime.
\newblock {\em IEEE Trans.\ on Inf.\ Th.}, 59(5):2545--2575, 2013.

\bibitem{Csi80}
I.~Csisz\'ar.
\newblock Joint source-channel error exponent.
\newblock {\em Problems of Control and Information Theory}, 9:315--328, 1980.

\bibitem{Nom13}
R.~Nomura and T.~S. Han.
\newblock Second-order {S}lepian-{W}olf coding theorems for non-mixed and mixed
  sources.
\newblock In {\em Int. Symp. Inf. Th.}, {Istanbul, Turkey}, 2013.
\newblock {\tt arXiv:1207.2505 [cs.IT]}.

\bibitem{verdu14}
S.~Verd\'u and I.~Kontoyiannis.
\newblock Optimal lossless data compression: Non-asymptotics and asymptotics.
\newblock {\em IEEE Trans. on Inf. Th.}, 60(2):777--795, Feb 2014.

\bibitem{cover75}
T.~M. Cover.
\newblock A proof of the data compression theorem of {Slepian and Wolf} for
  ergodic sources.
\newblock {\em IEEE Trans. Inf. Th.}, 21(3):226--228, March 1975.

\bibitem{miyake}
S.~Miyake and F.~Kanaya.
\newblock Coding theorems on correlated general sources.
\newblock {\em IEICE Trans.\ on Fundamentals of Electronics, Communications and
  Computer}, E78-A(9):1063--70, 1995.

\bibitem{Han10}
T.~S. Han.
\newblock {\em Information-Spectrum Methods in Information Theory}.
\newblock Springer Berlin Heidelberg, Feb 2003.

\bibitem{Pol13}
Y.~Polyanskiy.
\newblock Saddle point in the minimax converse for channel coding.
\newblock {\em IEEE Trans. on Inf. Th.}, 59:2576--2595, May 2013.

\bibitem{wang11}
D.~Wang, A.~Ingber, and Y.~Kochman.
\newblock The dispersion of joint source-channel coding.
\newblock In {\em Allerton Conference}, 2011.
\newblock {arXiv:1109.6310}.

\bibitem{feller}
W.~Feller.
\newblock {\em An Introduction to Probability Theory and Its Applications}.
\newblock John Wiley and Sons, 2nd edition, 1971.

\bibitem{Wang09}
L.~Wang, R.~Colbeck, and R.~Renner.
\newblock Simple channel coding bounds.
\newblock In {\em Intl.\ Symp.\ Inf.\ Th.}, Seoul, South Korea, 2009.

\bibitem{WangRenner}
L.~Wang and R.~Renner.
\newblock One-shot classical-quantum capacity and hypothesis testing.
\newblock {\em {Physical Review Letters}}, 108:200501, May 2012.

\bibitem{Dup12}
F.~Dupuis, L.~Kraemer, P.~Faist, J.~M. Renes, and R.~Renner.
\newblock Generalized entropies.
\newblock In {\em Proceedings of the XVIIth International Congress on
  Mathematical Physics}, 2012.

\bibitem{CS93}
N.~R. Chaganty and J.~Sethuraman.
\newblock Strong large deviation and local limit theorems.
\newblock {\em Annals of Probability}, 21(3):1671--1690, 1993.

\end{thebibliography}
